\documentclass[pra,superscriptaddress,floatfix,twocolumn,longbibliography]{revtex4-1}
\usepackage{graphicx}
\usepackage{wasysym}
\usepackage{color}
\def\newr{\color{black}}

\def\lsco{La$_{2-x}$Sr$_x$CuO$_4$}
\def\lbco{La$_{2-x}$Ba$_x$CuO$_4$}

\def\degrees{$^\circ$C}

\begin{document}

\title{Growth and structural characterization of large superconducting crystals of La$_{2-x}$Ca$_{1+x}$Cu$_2$O$_{6}$}

\author{J. A. Schneeloch}
\thanks{Present address: Department of Physics,
University of Virginia, Charlottesville, Virginia 22904, USA}
\affiliation{Condensed Matter Physics \&\ Materials Science Division, Brookhaven National Laboratory, Upton, NY 11973-5000, USA}
\author{Z. Guguchia}
\affiliation{Department of Physics, Columbia University, New York, New York 10027, USA}
\author{M. B. Stone}
\author{Wei Tian}
\affiliation{Quantum Condensed Matter Division, Oak Ridge National Laboratory, Oak Ridge, Tennessee 37831, USA}
\author{Ruidan Zhong}
\thanks{Present address: Department of Chemistry, Princeton University,
Princeton, New Jersey 08544, USA}
\affiliation{Condensed Matter Physics \&\ Materials Science Division, Brookhaven National Laboratory, Upton, NY 11973-5000, USA}
\affiliation{Materials Science and Engineering Department, Stony Brook University, Stony Brook, NY 11794, USA}
\author{K. M. Mohanty}
\author{Guangyong Xu}
\thanks{Present address: 
NIST Center for Neutron Research, National Institute of Standards and Technology, Gaithersburg, Maryland 20899, USA}
\author{G. D. Gu}
\author{J. M. Tranquada}
\email{jtran@bnl.gov}
\affiliation{Condensed Matter Physics \&\ Materials Science Division, Brookhaven National Laboratory, Upton, NY 11973-5000, USA}
\date{\today}
\begin{abstract}
Large crystals of La$_{2-x}$Ca$_{1+x}$Cu$_2$O$_{6}$ (La-Ca-2126) with $x=0.10$ and 0.15 have been grown and converted to bulk superconductors by high-pressure oxygen annealing. The superconducting transition temperature, $T_c$, is as high as 55~K; this can be raised to 60~K by post-annealing in air.  Here we present structural and magnetic characterizations of these crystals using neutron scattering and muon spin rotation techniques.  While the as-grown, non-superconducting crystals are single phase, we find that the superconducting crystals contain 3 phases forming coherent domains stacked along the $c$ axis: the dominant La-Ca-2126 phase, very thin (1.5 unit-cell) intergrowths of La$_2$CuO$_4$, and an antiferromagnetic La$_8$Cu$_8$O$_{20}$ phase.  We propose that the formation and segregation of the latter phases increases the Ca concentration of the La-Ca-2126, thus providing the hole-doping that supports superconductivity.
\end{abstract}
\maketitle

\section{Introduction}

The variety of cuprate compounds that exhibit superconductivity is remarkable; however, the set of cuprate families for which large superconducting crystals can readily be grown is much more limited.  Such large crystals are needed for neutron scattering studies of the magnetic excitations, where it is of interest to establish universal behaviors across multiple families.  While suitable crystals of La$_{2-x}$Sr$_x$CaCu$_2$O$_6$ (La-Sr-2126) \cite{ulri02} as well as La$_{2-x}$Ca$_{1+x}$Cu$_2$O$_6$ (La-Ca-2126)  \cite{wang03} have been grown previously and studied by neutron diffraction \cite{ulri02,huck05}, those samples exhibited little to no superconductivity.  Obtaining crystals with a large superconducting volume fraction has been a challenge, because synthesis of superconducting samples requires annealing in high-pressure oxygen \cite{cava90b}.  We have finally been able to achieve this last step, and in this paper we describe the synthesis conditions and magnetic and structural characterizations of the resulting crystals.

The La-Sr/Ca-2126 system is certainly not new.   The first report of La$_{2-x}$$A_{1+x}$Cu$_2$O$_{6-x/2}$ ($A = $Ca, Sr) was made by Raveau and coworkers \cite{nguy80} almost four decades ago.  Following the discovery of superconductivity in \lbco\ \cite{bedn86}, Torrance {\it et al.}\ \cite{torr88} reported on the metallic, but non-superconducting, character of La$_2$SrCu$_2$O$_{6.2}$, followed by crystallographic studies of La-Ca-2126 \cite{izum89} and La$_2$SrCu$_2$O$_6$ \cite{caig90}.  It was not long before Cava {\it et al.}\ \cite{cava90b} announced the discovery of superconductivity with a transition temperature, $T_c$, of $\sim60$~K in La-Sr-2126; the key here was to anneal in high-pressure oxygen.  Numerous studies of synthesis conditions and superconductivity in La-Sr-2126 \cite{saku91,liu91} and La-Ca-2126 \cite{fuer90,kino90,kino92a,kino92b} quickly followed.

Millimeter-size crystals of La-Ca-2126 were initially grown from a CuO flux; these were rendered superconducting, with $T_c$ as high as 40~K by annealing in an O$_2$ partial pressure of 300 atm at 1080$^\circ$C \cite{ishi91}.  Larger crystals (4 mm $\phi \times 30$ mm) were grown by the travelling-solvent floating-zone (TSFZ) method, where annealing in 400 atm of O$_2$ at 1080$^\circ$C yielded $T_c\approx45$~K \cite{okuy94}.   One of us (GDG) was able to grow large crystals of La-Sr-2126 in 11 bar O$_2$, which showed onset $T_c$'s over 40~K, but with a very small superconducting volume fraction ($<7$\%\ as measured by magnetic shielding) \cite{gu06b}.

Here we focus on crystals of La$_{2-x}$Ca$_{1+x}$Cu$_2$O$_6$ with $x=0.10$ and 0.15.  Annealing in $\gtrsim0.11$~GPa (1100 atm) partial pressure of O$_2$ yields sharp superconducting transitions with $T_c$ up to 60~K (after post-annealing).  The magnetic shielding fraction is essentially 100\%; however, there are substantial volume fractions of two other phases present, as we will demonstrate.

The rest of the paper is organized as follows.  The next section describes the crystal growth, annealing treatments, and the characterization methods [magnetization, neutron diffraction, and muon-spin-rotation ($\mu$SR)].  In Sec.~III, we present and analyze the diffraction and $\mu$SR data, followed by a description of the post-annealing study in Sec.~IV.  The results are discussed in Sec.~V and summarized in Sec.~VI.  We will present studies of the spin fluctuations by inelastic neutron scattering and anisotropic resistivity as a function of magnetic field in separate papers.

\section{Experimental Methods}

Single crystals of La-Ca-2126 were grown by the TSFZ method \cite{gu06a}. High purity powders of La$_{2}$O$_{3}$, CaCO$_{3}$, and CuO (99.99\%) were mixed in their metal ratio, ground well, and then sintered in a crucible. This grind-sinter procedure was repeated three times in order to achieve a homogeneous mixture, before the final feed-rod sintering, performed with the rod hung vertically from a Pt wire in a box furnace at 1100\degrees\ for 72 hours in air. The Cu-rich solvent material with lower melting point was prepared through the same procedures, and then sintered at 950 \degrees\ for 48 hours. Single-crystal growth was performed in flowing oxygen gas ($P_{\rm O_{2}}=1$~atm) in the floating-zone furnace. During the crystal growth, the feed and seed rods rotated in opposite directions at 30 rpm, to stir the liquid zone, and simultaneously translated through the heating zone at a velocity of 0.4 mm/h. After growth, the resulting rod was cut into sections, with polished sections checked with an optical polarization microscope to identify regions of single-crystal domain, allowing large single crystals to be selected for the high-pressure annealing experiments. 

The as-grown crystals of La-Ca-2126 are non-superconducting. We induced superconductivity in large crystals with both $x=0.10$ and 0.15 by annealing under a high-pressure mixture of 20\% oxygen and 80\% argon in a hot isostatic press (HIP).  The annealing was performed at temperatures within 1130--1180\degrees\ and pressures within 0.55--0.69 GPa in two separate runs, with run A lasting 31 hours and run B lasting 8 days {\newr \footnote{{\newr Note that the annealing temperature was selected to test the performance of the HIP.  Typical annealing temperatures in previous studies have been somewhat lower, in the range of 970$^\circ$C to 1080$^\circ$C \cite{cava90b,kino90,ishi91}.}}}.

\begin{figure}[t]
\centerline{\includegraphics[width=1.0\columnwidth]{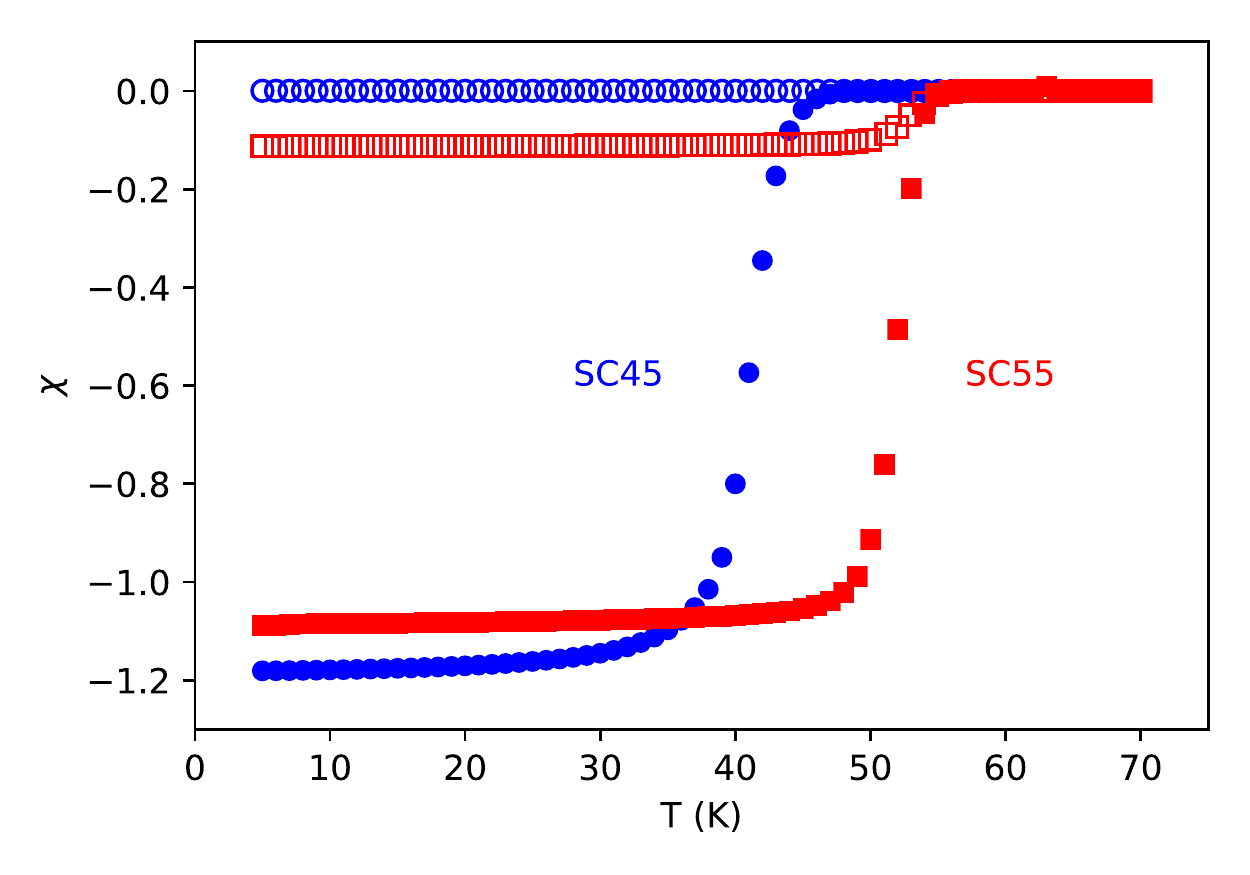}}
\caption{Volume magnetic susceptibility for samples SC45 (blue circles) and SC55 (red squares).  Open symbols: Meissner fraction (measured on field cooling); filled symbols: shielding fraction (zero-field cooling).}
\label{fg:susc1} 
\end{figure}

Magnetization measurements were performed using a commercial SQUID (superconducting quantum interference device) magnetometer. The magnetic susceptibility $\chi$ was calculated assuming a density of 6.244 g/cm$^{3}$, which was calculated from the lattice constants listed in Ref.\ \cite{huck05} assuming a composition of La$_{1.9}$Ca$_{1.1}$Cu$_{2}$O$_{6}$.   The susceptibility of two of the annealed crystals measured in a field of 1 mT (with uncertain crystal orientation) is shown in Fig.~\ref{fg:susc1}.  While the Meissner fraction, measured while field cooling, is small, the shielding fraction, obtained after zero-field cooling, is large.  

Initial neutron scattering experiments were performed on the SEQUOIA time-of-flight spectrometer at the Spallation Neutron Source, Oak Ridge National Laboratory \cite{tof_sns14}. Three crystals were used for neutron scattering, which we label NSC, SC45, and SC55. NSC is an as-grown, non-superconducting crystal with $x=0.10$ and mass 7.5g. SC45 is a 6.3~g crystal of $x=0.10$ annealed in run A.  SC55 is a 7.4-g crystal of $x=0.15$ annealed in run B.    As one can see from Fig.~\ref{fg:susc1}, SC45 has an onset of diamagnetism at $T_c=45$~K, while SC55 has $T_c=55$~K.  

While the inelastic scattering results will be reported separately, the data in the elastic channel provided interesting clues as to the structural phases in the sample, motivating further measurements.
Neutron diffraction measurements were then performed on triple-axis spectrometers HB-1A and HB-1 at the High Flux Isotope Reactor, Oak Ridge National Laboratory.  Sample SC55 was studied at HB-1A, where the incident energy was $E_i=14.6$~meV and horizontal collimations of $40'$-$40'$-S-$40'$-$80'$ were used.  The elastic scans of SC45 were done on a 1.2-g piece at HB-1 with $E_i=13.5$~meV and horizontal collimations $48'$-$40'$-S-$40'$-$120'$.  

\begin{figure}[b]
\centerline{\includegraphics[width=0.9\columnwidth]{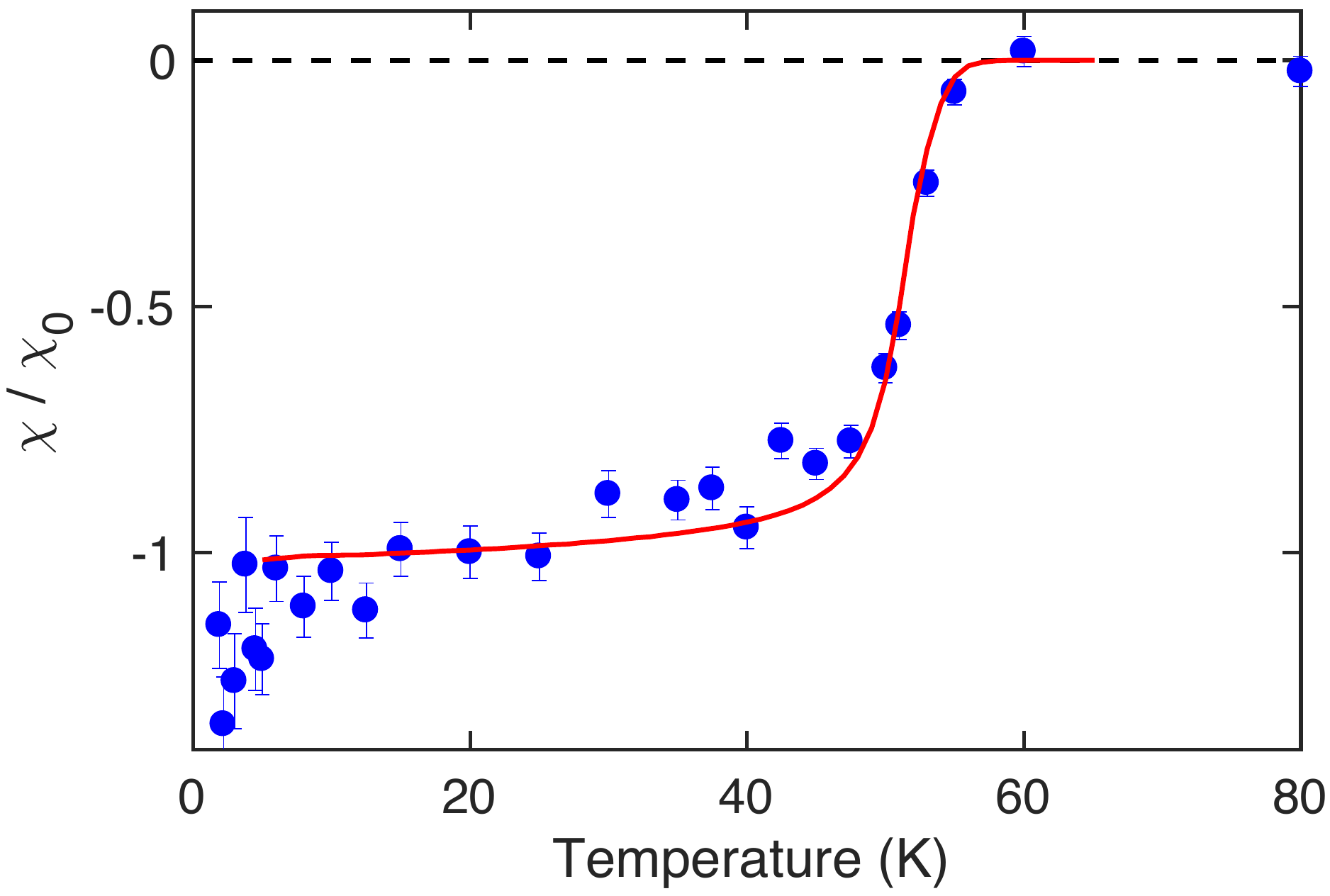}}
\caption{Normalized susceptibility data for pieces of the SC55$\mu$ sample.  Red line: zero-field-cooled bulk susceptibility; blue circles: diamagnetic shift of the internal field observed by muons in a field of 30~mT.}
\label{fg:susc2} 
\end{figure}

\begin{figure*}[t]
\centerline{\includegraphics[width=1.8\columnwidth]{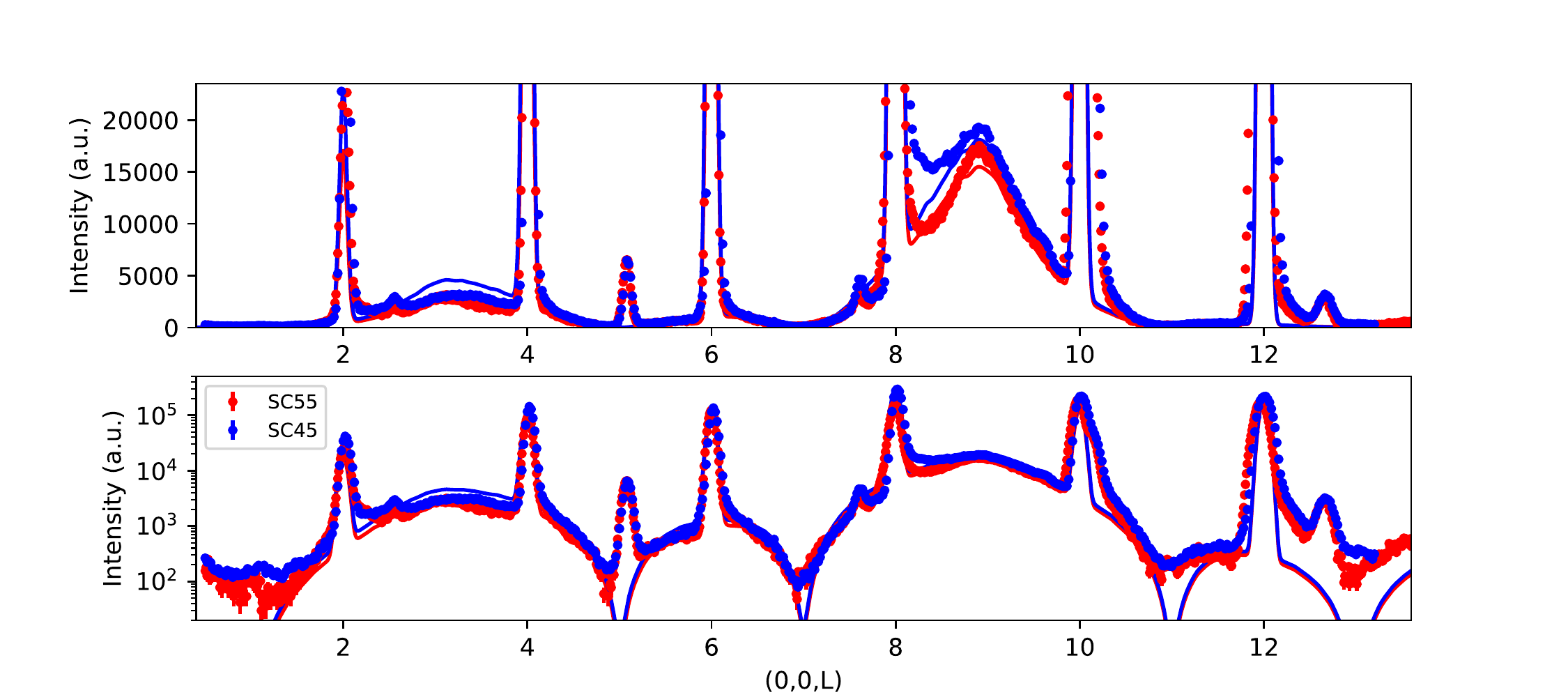}}
\caption{Neutron diffraction intensities measured along ${\bf Q}=(0,0,L)$ for the SC55 (red circles, measured on HB-1A) and SC45 (blue circles, measured on HB-1) crystals at $T=4$~K.  Intensity scale is in arbitrary units; data for the two samples have been normalized at the (004) peak.  Counting time was $\sim5$~s/pt, with an attenuator in the beam (to avoid detector saturation).  Top (bottom) panel uses a linear (logarithmic) intensity scale.  The solid lines are model calcuations as described in the text.}
\label{fg:00L} 
\end{figure*}

Complementary $\mu$SR measurements were performed at the $\pi$M3 beam line of the Paul Scherrer Institut (Switzerland), using the general purpose instrument (GPS), on an NSC crystal and a piece of $x=0.10$ annealed in run B, which we will label SC55$\mu$.   Experiments were performed both with the sample in zero field (ZF), to test for a finite magnetic hyperfine field, and in a transverse field (TF) of 3 mT, to probe the paramagnetic fraction in the normal state, or 30 mT, to probe the superconducting state.  The $\mu$SR spectra were analyzed in the time domain  using least-squares optimization routines from the {\tt musrfit} software suite \cite{sute12}.

Figure~\ref{fg:susc2} shows measurements on the SC55$\mu$ sample, comparing the normalized diamagnetic response obtained by weak-TF $\mu$SR and from a bulk susceptibility measurement.  Both measurements are quite consistent with $T_c\approx55$~K.

\section{Results and Analysis}
\label{sc:results}

A number of studies, largely on polycrystalline samples of La-Sr/Ca-2126, have demonstrated that high-pressure annealing can lead to secondary phases \cite{liu91,saku93,hu14b}.  This behavior can depend on the concentration of Ca or Sr.  We have chosen to focus on $x=0.10$ and 0.15 because previous work \cite{kino90} has indicated that these concentrations span the composition range for which homogeneous, single-phase samples can be prepared with the available oxygen partial pressure during crystal growth.  The high-pressure annealing of the crystals, essential for achieving bulk superconductivity, can result in some inhomogeneity. To identify specific phases and relative volumes of secondary phases in our superconducting crystals, we have combined neutron diffraction and $\mu$SR measurements.

\subsection{La-Ca-2126 and La-214}

In an earlier study of a high-pressure annealed La-Ca-2126 with $x=0.10$, imaging with transmission electron microscopy demonstrated the presence of intergrowth-like thin layers of La$_{2-x}$Ca$_x$CuO$_4$ (La-214) within the La-Ca-2126 matrix \cite{hu14b}.  (Note that such layers were not observed in an as-grown crystal.)  From neutron diffraction measurements, such as those shown in Fig.~\ref{fg:00L}, we again find evidence for both phases.  The sharp Bragg peaks at $(00L)$ with $L$ an even integer come from the La-Ca-2126 domains, while the La-214 phase shows up as broad diffuse scattering  with prominent  peaks at $L\sim3.2$ and 8.9.  (Other peaks, such as those at $L\approx2.5$, 5.1, 7.7, 10.2, and 12.8, come from a third phase that will be discussed in the following subsection.)

To model these data, we created a random stacking along the $c$ axis of two structural units: 0.5 unit cell of La$_2$CaCu$_2$O$_6$ and 1.5 unit cells of La$_2$CuO$_4$.  These units are terminated by LaO layers, and there is a shift of $(0.5,0.5,0)$ applied between neighboring units.  To calculate the structure factor, structural parameters for La-Ca-2126 were taken from \cite{ulri02} and for La-214 from \cite{jorg88}, where we take the space group to be $I4/mmm$ in both cases.  The lattice parameters used for La-Ca-2126 are $a=3.83$~\AA\ and $c=19.37$~\AA.  For La-214, we started with $c=13.1$~\AA, corresponding to the undoped bulk compound; however, we found that the fit was considerably improved by using $c_{214}=\frac23 c_{2126} = 12.91$~\AA.  

If we allow that the La-214 layers have their $a$ lattice parameter epitaxially-constrained to match that of La-Ca-2126, an expansion of 1.007, then the reduction in $c$ corresponds to a uniaxial strain of the unit cell that does not change the unit cell volume.  Given that Ca solubility in La$_2$CuO$_4$ is limited and its presence tends to change the lattice parameters in directions opposite to the apparent strain \cite{mood92}, it seems likely that the La-214 phase contains minimal Ca ($x\lesssim0.1$).

The model calculations shown in Fig.~\ref{fg:00L} correspond to the square of the structure factor calculated with 29,500 La-Ca-2126 units and 3,500 La-214 units.  These parameters are a compromise that is close to the best fit for both samples.  The intensity has been multiplied by an $L$-dependent correction for spectrometer resolution volume; the only differences between the two calculated curves are due to the differences in spectrometer configurations for the measurements of the two samples.  We find that this model gives a very good description of the measured scattering.  Taking into account the relative thicknesses of the structural units, the calculation shown corresponds to an 80.8\%\ volume fraction of La-Ca-2126 and 19.2\%\ of La-214.  The actual best fits to each data set correspond to a La-214 volume fraction of 17.7\%\ for SC45 and 19.9\%\ for SC55.  The difference between these values is comparable to the uncertainty, but it suggests that the volume of La-214 increases with annealing time and $T_c$.

\subsection{La-8-8-20}

\begin{figure}[b]
\centerline{\includegraphics[width=1.0\columnwidth]{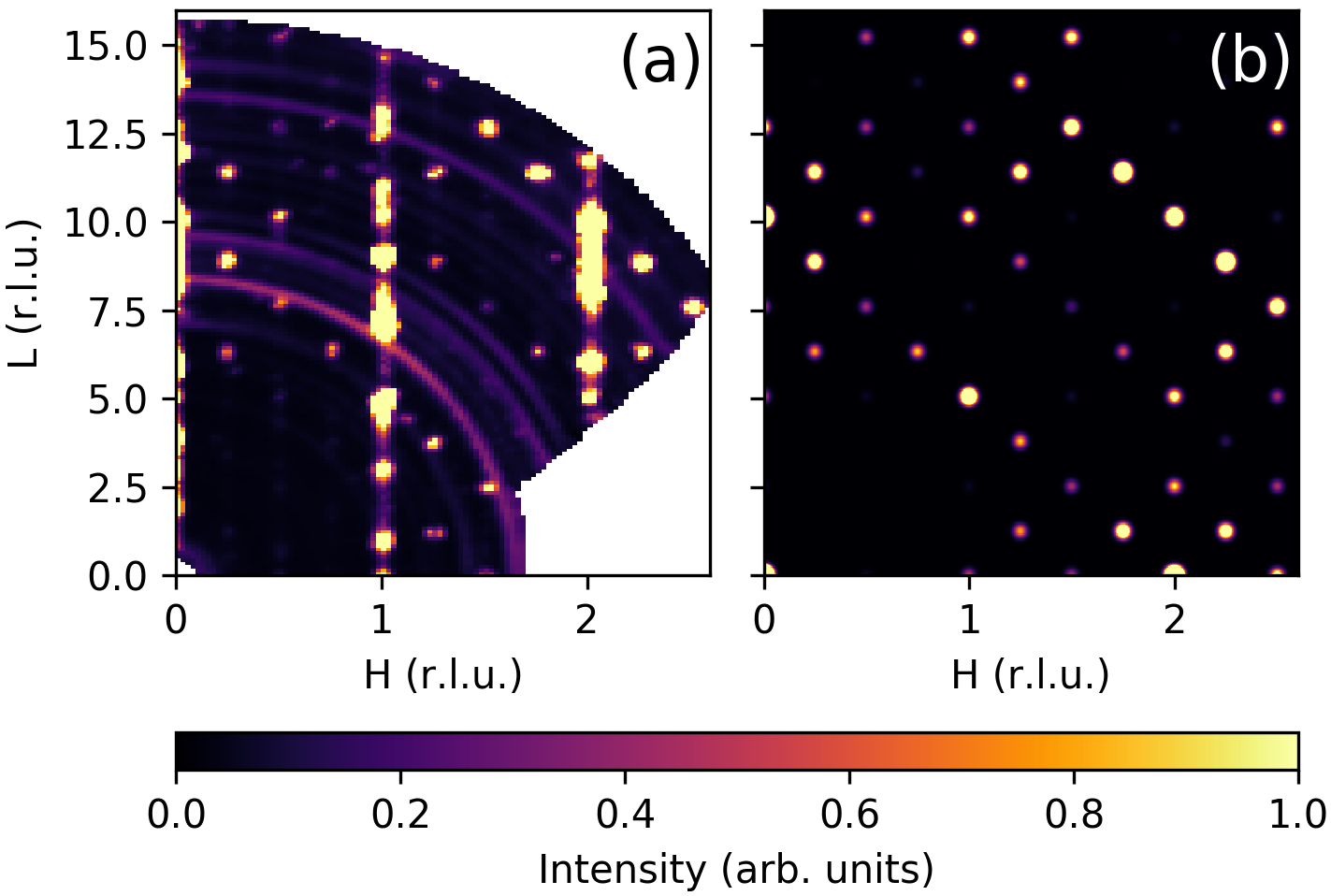}}
\caption{(a) Elastic scattering obtained at SEQUOIA on the SC55 sample at $T=4$~K.  Rings of scattering correspond to powder diffraction from the Al sample holder and from the sample itself.  (b) Calculated diffraction pattern for the La-8-8-20 phase using the structural parameters from \cite{erra88}.}
\label{fg:tof} 
\end{figure}

Another phase that has been identified in La-Sr-2126 samples is La$_{8-x}$Sr$_x$Cu$_8$O$_{20}$ (La-8-8-20) \cite{liu91,saku93,erra88}.  This phase was first reported in 1987 as La$_5$SrCu$_6$O$_{15}$ \cite{toku87,torr88}; the proper formula per unit cell was later determined in a neutron powder diffraction study \cite{erra88}.    Given the absence of Sr in our samples and evidence (discussed in the next subsection) that the relevant phase is an antiferromagnetic insulator, we believe that our case corresponds to $x=0$.  The phase is essentially a version of the perovskite LaCuO$_{3-\delta}$ with an ordered arrangement of oxygen vacancies.  Taking $a_0$ as the average Cu-O-Cu distance, the unit cell is tetragonal with $a'\approx2\sqrt{2}a_0$ and $c'\approx a_0$.  Extrapolating reported lattice parameters for finite $x$ to $x=0$ gives $a'\approx10.89$~\AA\ and $c'\approx3.85$~\AA, with $a_0\approx3.85$~\AA.

We can fully explain the extra $(00L)$ peaks in Fig.~\ref{fg:00L} if the La-8-8-20 phase is oriented coherently with the La-Ca-2126 phase, such that a [110] axis of the former is parallel to the [001] axis of the latter.   This identification also allows us to explain an array of superlattice peaks observed in the elastic channel of time-of-flight measurements performed at SEQUOIA.  An example of peaks in the $(H0L)$ zone for the SC55 sample is shown in Fig.~\ref{fg:tof}, together with peak positions and intensities calculated from the reported structural parameters \cite{erra88}.  Based on the analysis of Bragg peak intensities, we estimate that the La-8-8-20 phase corresponds to $\sim15\%$ of the sample volume.   The presence of this third phase then renormalizes the fractions of the other phases to 69\%\ La-Ca-2126 and 16\%\ La-214.

\subsection{Antiferromagnetism}

\begin{figure}[b]
\centerline{\includegraphics[width=0.9\columnwidth]{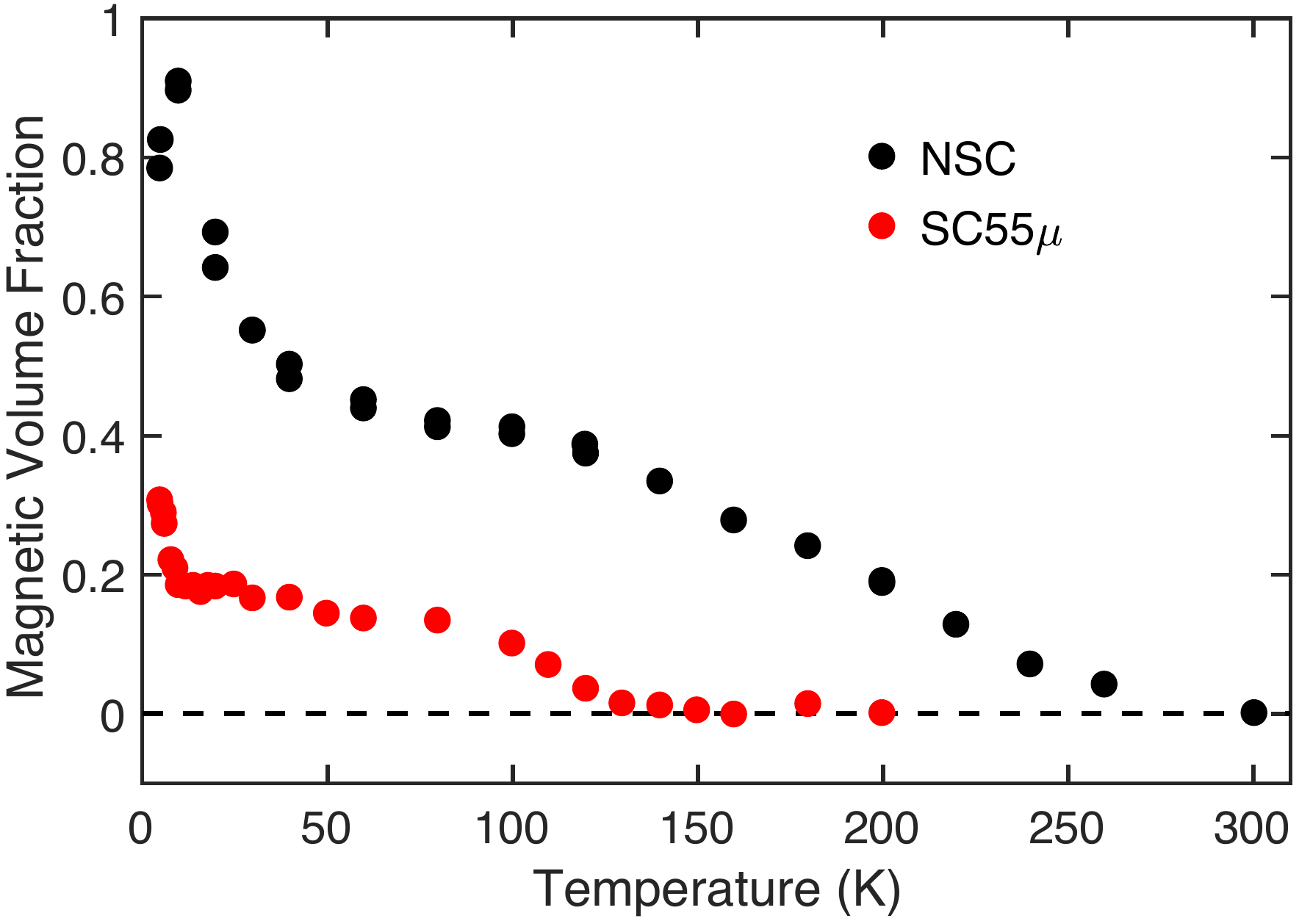}}
\caption{Magnetic volume fraction in the NSC and SC55$\mu$ samples as determined from $\mu$SR measured in a weak transverse field of 3 mT. }
\label{fg:mvol} 
\end{figure}

Zero-field $\mu$SR data provide clear evidence for long-range antiferromagnetic ordering in the NSC sample, with two distinct precession frequencies in the $\mu$SR signal, corresponding (at low temperature) to the local magnetic fields 37.7(1)~mT (70\%\ of the signal) and 102.6(3)~mT (30\%\ of the signal).  The hyperfine fields (sampled at only a few temperatures) grow substantially on cooling, especially between 100~K and 5~K.  Similar measurements on the SC55$\mu$ sample provide evidence for ordering, but without any coherent precession signal; instead, a fast decaying ${\mu}$SR signal is found, which could be due to a wide distribution of static fields.  Figure~\ref{fg:mvol} shows the temperature-dependent magnetic volume fractions for the two samples determined from weak TF $\mu$SR.  Both samples show a significant enhancement of the magnetic volume fraction at low temperature.

The rise in magnetic volume and hyperfine fields at low $T$ in the NSC sample is reminiscent of related behavior in lightly-doped \lsco, with $0<x<0.02$.  There, the Sr doping caused a reduction of the N\'eel temperature and of the hyperfine field observed by $\mu$SR \cite{bors95} and La nuclear quadrupole resonance \cite{chou93}, but the full hyperfine field of the $x=0$ system was recovered on cooling to low temperature, with the recovery beginning below 30~K.  Neutron scattering studies later showed that the low-temperature recovery corresponds to phase separation of the doped holes into patches of diagonal spin stripes \cite{mats02}.  The segregation of the holes allows the hyperfine field in the commensurate antiferromagnetic regions to achieve its full strength.

Returning to the NSC sample, previous neutron scattering measurements on a similar as-grown La-Ca-2126 crystal with $x=0.1$ found a slight downturn in the magnetic Bragg peak intensity below 25~K; however, there was a concomitant increase in the elastic diffuse magnetic scattering from CuO$_2$ bilayers \cite{huck05}.  It seems likely that these changes reflect some type of segregation of a low density of doped holes, which is associated with the rise in the ordered magnetic volume fraction below 50~K, as indicated in Fig.~\ref{fg:mvol}.

\begin{figure}[t]
\centerline{\includegraphics[width=1.0\columnwidth]{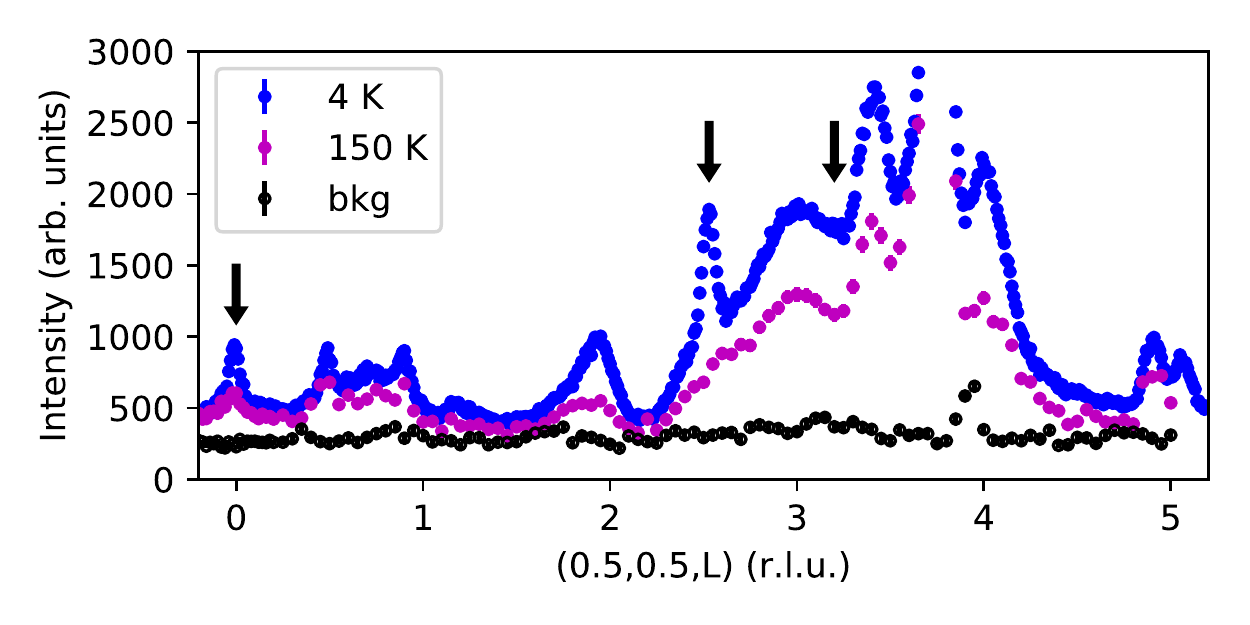}}
\caption{Neutron diffraction scans along ${\bf Q}=(\frac12,\frac12,L)$ measured at $T=4$~K (blue points) and 150~K (violet points) on HB-1A.  Background data (black points) measured at ${\bf Q}=(0.45,0.45,L)$ and $T=4$.  Three of the $L$ values at which the temperature dependence was followed in detail are indicated by arrows.}
\label{fg:magL} 
\end{figure}

For the SC55$\mu$ sample, a somewhat different interpretation is required.  Here we believe that none of the magnetic order is in the La-Ca-2126 phase.  The neutron diffraction data discussed below indicate that the magnetic order that onsets at $\sim130$~K is associated with the La-8-8-20 phase.  The upturn below 10~K is reminiscent of the spin-glass behavior observed in lightly-doped \lsco\ \cite{nied98}, and we propose that it is associated with the thin La-214 layers.  {\newr Based on the upturn temperature, we infer doping by Ca to a hole concentration of less than 0.1.  Assuming that the behavior is similar to that of the bulk, oxygen doping could not explain this behavior since phase separation tends to yield an antiferromagnetic phase with a N\'eel temperature of $\sim260$~K plus a superconducting phase \cite{well97}.}

\begin{figure}[t]
\centerline{\includegraphics[width=1.0\columnwidth]{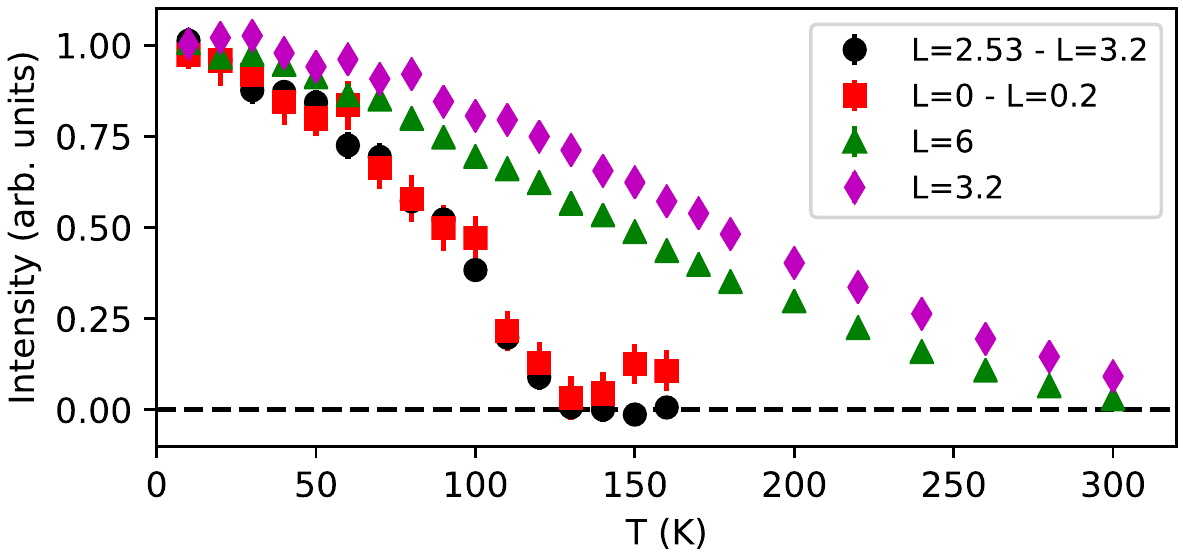}}
\caption{Temperature dependence of intensities for ${\bf Q}=(0.5,0.5,L)$ at several values of $L$, after appropriate background subtraction.  Red squares: $L=0$; black circles: $L=2.53$; violet diamonds: $L=3.2$; green triangles: $L=6$.}
\label{fg:magT} 
\end{figure}

To characterize antiferromagnetic order in SC55, we performed diffraction scans along 
$(\frac12,\frac12,L)$, as shown in Fig.~\ref{fg:magL}.  Much of the scattering is nuclear, such as the broad diffuse scattering peaked at $L\sim3.5$, which is partially reduced on warming from 4~K to 150~K.  In contrast, the relatively sharp peaks at $L=0$ and $L=2.53$ have largely disappeared by 150~K.  In Fig.~\ref{fg:magT}, we compare the temperature dependences of these peaks with diffuse scattering at $L=3.2$ and a strong La2126 structural superlattice peak at $L=6$.  The intensities of the former peaks drop to zero at $\sim130$~K, consistent with the magnetic ordering temperature from the $\mu$SR measurements shown in Fig.~\ref{fg:mvol}, while the diffuse and structural superlattice intensities only approach zero near 300~K.

The fact that the $L=0$ and 2.53 peak intensities show the same temperature dependence suggests that they come from the same phase.  The $L$ value of 2.53 is uniquely consistent with the La-8-8-20 phase, corresponding to a commensurate reflection.  [In terms of the cubic perovskite cell with $a=3.85$~\AA, it corresponds to $(\frac12,\frac12,\frac12)$.] Hence, it appears that the La-8-8-20 phase is magnetic, and the corresponding magnetic volume fraction for $T\gtrsim50$~K is consistent with our diffraction estimate of about 15\%\ sample volume.  The upturn below 10~K is compatible with the net 17\%\ volume of La-214 layers estimated from the analysis of the $(00L)$ neutron scattering data.

\section{Post-annealing in air}

As a test, a 67-mg piece of the SC55 sample was annealed in air at several different temperatures.   As a reference on mass change, a 92-mg as-grown, non-superconducting crystal with $x=0.15$ was included in the annealing.  Sample masses were measured before and after each annealing, each annealing was for 24h, and the temperature dependence of the magnetic susceptibility was measured (zero-field cooling) after each anneal.

\begin{figure}[t]
\centerline{\includegraphics[width=0.65\columnwidth]{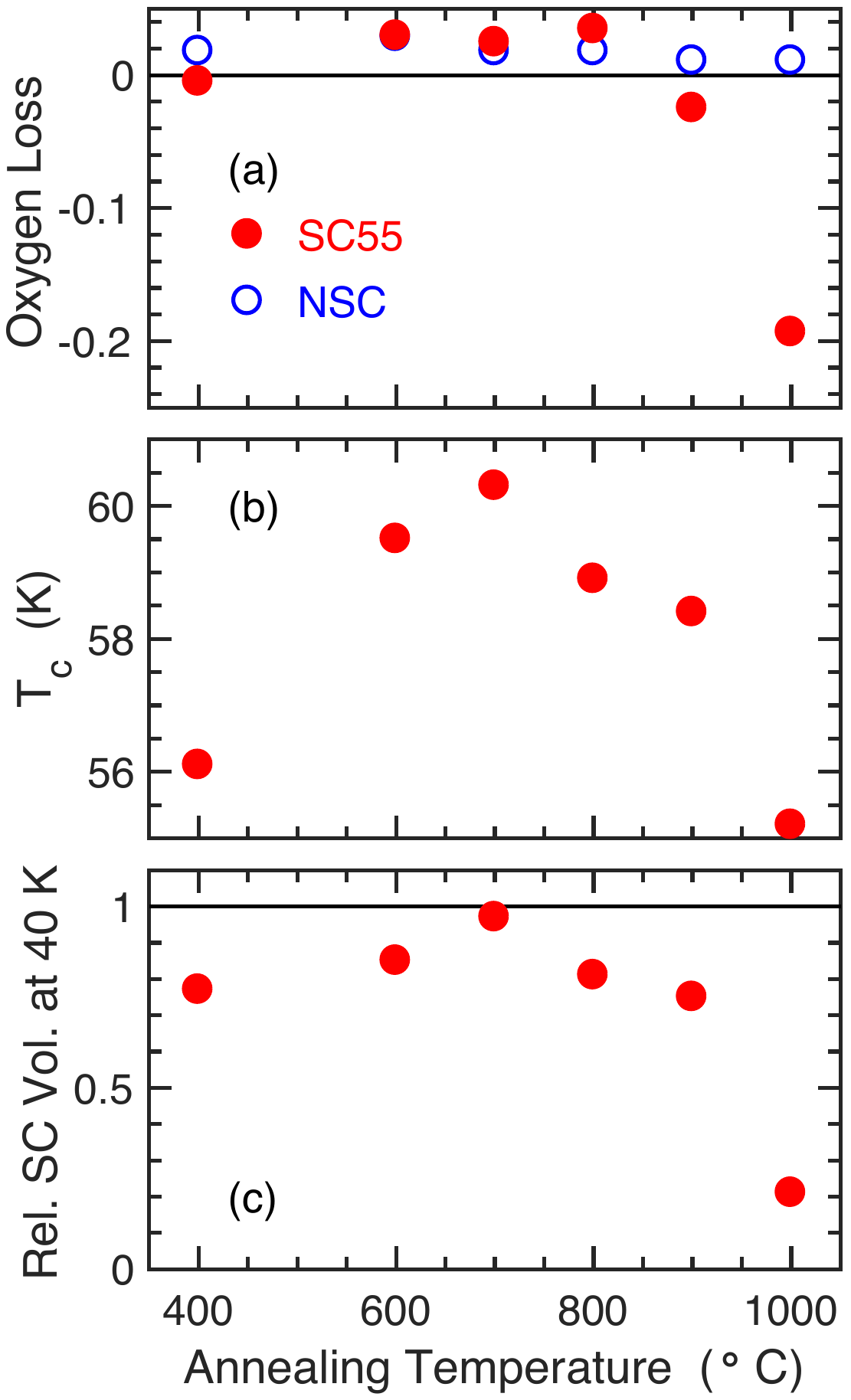}}
\caption{Results obtained from a series of 24h annealings in air at steadily increasing temperatures.  (a) Cumulative change in oxygen content, normalized to the La-Ca-2126 formula unit, estimated from mass changes for a piece of SC55 and an as-grown, non-superconducting piece of La-Ca-2126 with $x=0.15$.  (b) Resulting $T_c$ after each annealing, determined from ZFC susceptibility at $\chi=-0.1$.  (c)  Superconducting (shielding) volume at 40~K, relative to that of the 700$^\circ$C sample at 5~K, from the ZFC susceptibility data.}
\label{fg:anneal} 
\end{figure}

The results are summarized in Fig.~\ref{fg:anneal}.  We make the assumption that any mass change corresponds to a variation in the oxygen content, and Fig.~\ref{fg:anneal}(a) shows the cumulative change in oxygen content, normalized to the formula unit.  The first meaningful change in oxygen content occurs from annealing at 900$^\circ$C, and it grows substantially at 1000$^\circ$C.  The superconducting $T_c$ actually rises to a maximum {\newr (60~K)} after annealing at 700$^\circ$C, before decreasing slightly at higher annealing temperatures, as shown in Fig.~\ref{fg:anneal}(b).  The relative superconducting volume (shielding) fraction seems to show small changes that roughly correlate with $T_c$, but it plummets when the oxygen content drops, as shown in Fig.~\ref{fg:anneal}(c).

\section{Discussion}
\label{sc:discussion}

{\newr We have assumed that the superconductivity in our sample comes from the La-Ca-2126 phase, but we discovered that the annealed crystals also contain two other phases.  Could one of the latter phases be responsible for the superconductivity?  We believe that the combination of the neutron and $\mu$SR data make a convincing case that the La-8-80-20 phase must be antiferromagnetic with a N\'eel temperature of $\sim130$~K; such order is inconsistent with superconductivity.  La-214 can be superconducting; however, the highest reported $T_c$ due to Ca doping is just 34~K \cite{dabr93}.  Oxygen doping of La-214 can raise $T_c$ as high as 45~K \cite{blak98}; however, oxygen intercalation increases the $c$ lattice parameter \cite{rada93}, which is inconsistent with the decreased $c$ parameter inferred for our La-214 intergrowths.  Hence, we believe that the main source of the superconductivity in our crystals must be the La-Ca-2126 phase, especially given the observed $T_c$ of 55~K. }

An unresolved issue for La-Ca-2126 concerns the mechanism for doping sufficient holes into the CuO$_2$ planes to achieve the observed superconducting transition temperatures.  Suppose we write the chemical formula as La$_{2-x}$Ca$_{1+x}$Cu$_2$O$_{6+\delta}$, to allow for oxygen interstitials or vacancies.  Then the effective hole concentration $p$ per Cu would be $p = x/2 + \delta$, and the maximum $T_c$ in a given cuprate family, corresponding to ``optimal'' doping, typically occurs for $p_{\rm opt}\sim0.16$.  In principle, we could achieve $p_{\rm opt}$ by using sufficient Ca, adding excess oxygen, or using a combination of these components.   If we consider the composition originally studied by Cava {\it et al.} \cite{cava90b,cava90c}, La$_{1.6}$Sr$_{0.4}$CaCu$_2$O$_{6+\delta}$ with $\delta\sim-0.07$, application of our formula gives $p = 0.13$, which is in the vicinity of $p_{\rm opt}$.  There, the doping by excess alkaline-earth ions was partially offset by oxygen vacancies.

Next, consider our samples and assume, for the moment, that $\delta=0$ and that we have a single phase in the superconducting samples.  For $x=0.10$, we would get $p=0.05$, which would generally be too low to realize superconductivity; increasing $x$ to 0.15 only just crosses the anticipated threshold into the superconducting regime.  One might imagine that annealing in high-pressure oxygen would increase the oxygen content of the sample, raising $p$ towards $p_{\rm opt}$.   A few studies have used neutron powder diffraction to analyze the change in oxygen occupancy due to such annealing, including consideration of vacancies at regular lattice sites and  additional atoms at an extra oxygen site within the CuO$_2$ bilayers \cite{saku91,kino92a,shak93}; however, the associated change in hole concentration would only appear to be $\Delta p\sim0.03$ (with a large spread in results), and this would not be sufficient to shift the $T_c$ of our samples from zero to 60~K.

Of course, our crystals have extra phases in them, and we believe that the formation of these secondary phases is essential to the doping process.  The key point is that the La-214 and La-8-8-20 phases contain very little Ca compared to the La-Ca-2126 phase.  We start with a nominally uniform composition of La-Ca-2126 with an excess Ca concentration of $x$.  Removing quantities of the elements La, Cu, and O to form La-214 and La-8-8-20 phases will increase the Ca concentration to $x'$ in the remaining La-Ca-2126 phase.  (A small amount of Ca in the La-214 phase has only a small impact on $x'$.)  Using the estimated phase fractions, a simple analysis yields $x' \sim 0.6$, which, in the absence of oxygen vacancies, would predict $p\gg p_{\rm opt}$ {\newr \footnote{{\newr In principle, one could estimate $x'$ from a refinement of neutron diffraction data; however, we did not collect data suitable for such a refinement.  The triple-axis measurements covered only limited directions in reciprocal space.  The time-of-flight measurements covered a bigger range of reciprocal space; however, those measurements were aimed at inelastic scattering and the data were not collected in a fashion that would ensure reliable Bragg peak intensities.}}}.  {\newr Based on the fact that the $T_c$'s of our samples are comparable to the maximum values obtained in previous studies, we assume that the actual hole concentration cannot be more than slightly beyond optimal doping.}

In equilibrium, the relative volumes of the various phases should be controlled by the chemical potential.  Kinetics may limit whether we reach equilibrium.  The hole concentration in the La-Ca-2126 phase is also associated with a chemical potential.  As the volume fraction of the secondary phases increases, a constant chemical potential could cause oxygen vacancies to develop in the La-Ca-2126 phase to compensate for the growing Ca concentration.  This may keep our superconducting samples from reaching the over-doped regime.

This surprising doping mechanism also seems to be compatible with the observation that the final $T_c$ of a crystal does not seem to depend on the initial value of $x$ (for our narrow range of $x$), but only the length of time spent at high pressure.  We have observed that the volume fraction of La-214 is increased by a longer annealing time. Note that we have attempted to maintain the same temperature during annealing; we have not yet studied the impact of varying the annealing temperature in a controlled fashion.   If doping were controlled by introducing excess oxygen into La-Ca-2126, then we would expect that the oxygen content should be determined by the annealing temperature and pressure; changing the annealing time at a fixed temperature and pressure might impact the transition width, but it should not affect the onset $T_c$.  Instead, we see sharp transitions with $T_c$ values that depend on time alone.  Given the large inferred value of $x'$ in the La-Ca-2126 phase, it may be possible to attain a more modest value of $x'$ and achieve superconductivity, with a smaller volume fraction of secondary phases, by annealing at a somewhat lower temperature.  This will be tested in the future.

The impact of the post-annealing on $T_c$ will require further work to explain.  Does it involve a tuning of the doping through a change in concentration of the secondary phases, or possibly a relaxation of strain between epitaxial phases {\newr or ordering of oxygen interstitials/vacancies}?  Resolving this will require a careful {\newr structural analysis} 
of samples at various stages of post annealing.  

\section{Summary}
\label{sc:summary}

We have grown large crystals of La$_{2-x}$Ca$_{1+x}$Cu$_2$O$_6$ with $x=0.10$ and 0.15.  The crystals are non-superconducting as grown.  Superconductivity has been induced by annealing in $\sim0.12$~GPa partial pressure of O$_2$ at $T\sim 1150^\circ$C.  We have obtained crystals with sharp transitions at either $T_c=45$~K or 55~K, with $T_c$ growing with annealing time.

We have used neutron scattering and $\mu$SR to characterize the structural and magnetic phases in these crystals.  The as-grown crystals are essentially single phase, with significant antiferromagnetic order.  The superconducting crystals, in contrast, contain two secondary epitaxial phases, La$_{2-x}$Ca$_x$CuO$_4$ and La$_8$Cu$_8$O$_{20}$.  The La-8-8-20 phase exhibits antiferromagnetic order below 130~K, while the La-214 may develop spin-glass order below 10~K.   It appears that the formation of the secondary phases enhances the Ca concentration of the La-Ca-2126 phase, and it is the resulting change in hole-doping that induces the superconductivity.

\acknowledgments
 
Work at Brookhaven was supported by the Office of Basic Energy Sciences (BES), Division of Materials Sciences and Engineering, U.S. Department of Energy (DOE), through Contract No.\ DE-SC0012704.     Z.G. gratefully acknowledges financial support by the Swiss National Science Foundation (Early Postdoc Mobility SNF fellowship P2ZHP2-161980), and sincerely thanks Hubertus Luetkens and Alex Amato for support on the $\mu$SR experiments.
RZ was supported by the Center for Emergent Superconductivity, an Energy Frontier Research Center funded by BES.  The experiments at ORNL's SNS and HFIR were sponsored by the Scientific User Facilities Division, BES, U.S. DOE.  

\bibliography{lno,theory,neutrons}

\begin{thebibliography}{39}%
\makeatletter
\providecommand \@ifxundefined [1]{%
 \@ifx{#1\undefined}
}%
\providecommand \@ifnum [1]{%
 \ifnum #1\expandafter \@firstoftwo
 \else \expandafter \@secondoftwo
 \fi
}%
\providecommand \@ifx [1]{%
 \ifx #1\expandafter \@firstoftwo
 \else \expandafter \@secondoftwo
 \fi
}%
\providecommand \natexlab [1]{#1}%
\providecommand \enquote  [1]{``#1''}%
\providecommand \bibnamefont  [1]{#1}%
\providecommand \bibfnamefont [1]{#1}%
\providecommand \citenamefont [1]{#1}%
\providecommand \href@noop [0]{\@secondoftwo}%
\providecommand \href [0]{\begingroup \@sanitize@url \@href}%
\providecommand \@href[1]{\@@startlink{#1}\@@href}%
\providecommand \@@href[1]{\endgroup#1\@@endlink}%
\providecommand \@sanitize@url [0]{\catcode `\\12\catcode `\$12\catcode
  `\&12\catcode `\#12\catcode `\^12\catcode `\_12\catcode `\%12\relax}%
\providecommand \@@startlink[1]{}%
\providecommand \@@endlink[0]{}%
\providecommand \url  [0]{\begingroup\@sanitize@url \@url }%
\providecommand \@url [1]{\endgroup\@href {#1}{\urlprefix }}%
\providecommand \urlprefix  [0]{URL }%
\providecommand \Eprint [0]{\href }%
\providecommand \doibase [0]{http://dx.doi.org/}%
\providecommand \selectlanguage [0]{\@gobble}%
\providecommand \bibinfo  [0]{\@secondoftwo}%
\providecommand \bibfield  [0]{\@secondoftwo}%
\providecommand \translation [1]{[#1]}%
\providecommand \BibitemOpen [0]{}%
\providecommand \bibitemStop [0]{}%
\providecommand \bibitemNoStop [0]{.\EOS\space}%
\providecommand \EOS [0]{\spacefactor3000\relax}%
\providecommand \BibitemShut  [1]{\csname bibitem#1\endcsname}%
\let\auto@bib@innerbib\@empty
\bibitem [{\citenamefont {Ulrich}\ \emph {et~al.}(2002)\citenamefont {Ulrich},
  \citenamefont {Kondo}, \citenamefont {Reehuis}, \citenamefont {He},
  \citenamefont {Bernhard}, \citenamefont {Niedermayer}, \citenamefont
  {Bour\'ee}, \citenamefont {Bourges}, \citenamefont {Ohl}, \citenamefont
  {R\o{}nnow}, \citenamefont {Takagi},\ and\ \citenamefont {Keimer}}]{ulri02}%
  \BibitemOpen
  \bibfield  {author} {\bibinfo {author} {\bibfnamefont {C.}~\bibnamefont
  {Ulrich}}, \bibinfo {author} {\bibfnamefont {S.}~\bibnamefont {Kondo}},
  \bibinfo {author} {\bibfnamefont {M.}~\bibnamefont {Reehuis}}, \bibinfo
  {author} {\bibfnamefont {H.}~\bibnamefont {He}}, \bibinfo {author}
  {\bibfnamefont {C.}~\bibnamefont {Bernhard}}, \bibinfo {author}
  {\bibfnamefont {C.}~\bibnamefont {Niedermayer}}, \bibinfo {author}
  {\bibfnamefont {F.}~\bibnamefont {Bour\'ee}}, \bibinfo {author}
  {\bibfnamefont {P.}~\bibnamefont {Bourges}}, \bibinfo {author} {\bibfnamefont
  {M.}~\bibnamefont {Ohl}}, \bibinfo {author} {\bibfnamefont {H.~M.}\
  \bibnamefont {R\o{}nnow}}, \bibinfo {author} {\bibfnamefont {H.}~\bibnamefont
  {Takagi}}, \ and\ \bibinfo {author} {\bibfnamefont {B.}~\bibnamefont
  {Keimer}},\ }\bibfield  {title} {\enquote {\bibinfo {title} {{Structural and
  magnetic instabilities of
  ${\mathrm{La}}_{2\ensuremath{-}x}{\mathrm{Sr}}_{x}{\mathrm{CaCu}}_{2}{\mathrm{O}}_{6}$}},}\
  }\href {\doibase 10.1103/PhysRevB.65.220507} {\bibfield  {journal} {\bibinfo
  {journal} {Phys. Rev. B}\ }\textbf {\bibinfo {volume} {65}},\ \bibinfo
  {pages} {220507} (\bibinfo {year} {2002})}\BibitemShut {NoStop}%
\bibitem [{\citenamefont {Wang}\ \emph {et~al.}(2003)\citenamefont {Wang},
  \citenamefont {Zheng}, \citenamefont {Feng}, \citenamefont {Gu},
  \citenamefont {Homes}, \citenamefont {Tranquada}, \citenamefont {Gaulin},\
  and\ \citenamefont {Timusk}}]{wang03}%
  \BibitemOpen
  \bibfield  {author} {\bibinfo {author} {\bibfnamefont {N.~L.}\ \bibnamefont
  {Wang}}, \bibinfo {author} {\bibfnamefont {P.}~\bibnamefont {Zheng}},
  \bibinfo {author} {\bibfnamefont {T.}~\bibnamefont {Feng}}, \bibinfo {author}
  {\bibfnamefont {G.~D.}\ \bibnamefont {Gu}}, \bibinfo {author} {\bibfnamefont
  {C.~C.}\ \bibnamefont {Homes}}, \bibinfo {author} {\bibfnamefont {J.~M.}\
  \bibnamefont {Tranquada}}, \bibinfo {author} {\bibfnamefont {B.~D.}\
  \bibnamefont {Gaulin}}, \ and\ \bibinfo {author} {\bibfnamefont
  {T.}~\bibnamefont {Timusk}},\ }\bibfield  {title} {\enquote {\bibinfo {title}
  {{Infrared properties of
  ${\mathrm{La}}_{2\ensuremath{-}x}(\mathrm{C}\mathrm{a},\mathrm{S}\mathrm{r}{)}_{x}{\mathrm{CaCu}}_{2}{\mathrm{O}}_{6+\ensuremath{\delta}}$
  single crystals}},}\ }\href {\doibase 10.1103/PhysRevB.67.134526} {\bibfield
  {journal} {\bibinfo  {journal} {Phys. Rev. B}\ }\textbf {\bibinfo {volume}
  {67}},\ \bibinfo {pages} {134526} (\bibinfo {year} {2003})}\BibitemShut
  {NoStop}%
\bibitem [{\citenamefont {H\"ucker}\ \emph {et~al.}(2005)\citenamefont
  {H\"ucker}, \citenamefont {Kim}, \citenamefont {Gu}, \citenamefont
  {Tranquada}, \citenamefont {Gaulin},\ and\ \citenamefont {Lynn}}]{huck05}%
  \BibitemOpen
  \bibfield  {author} {\bibinfo {author} {\bibfnamefont {M.}~\bibnamefont
  {H\"ucker}}, \bibinfo {author} {\bibfnamefont {Young-June}\ \bibnamefont
  {Kim}}, \bibinfo {author} {\bibfnamefont {G.~D.}\ \bibnamefont {Gu}},
  \bibinfo {author} {\bibfnamefont {J.~M.}\ \bibnamefont {Tranquada}}, \bibinfo
  {author} {\bibfnamefont {B.~D.}\ \bibnamefont {Gaulin}}, \ and\ \bibinfo
  {author} {\bibfnamefont {J.~W.}\ \bibnamefont {Lynn}},\ }\bibfield  {title}
  {\enquote {\bibinfo {title} {{Neutron scattering study on
  ${\mathrm{La}}_{1.9}{\mathrm{Ca}}_{1.1}{\mathrm{Cu}}_{2}{\mathrm{O}}_{6+\ensuremath{\delta}}$
  and
  ${\mathrm{La}}_{1.85}{\mathrm{Sr}}_{0.15}\mathrm{Ca}{\mathrm{Cu}}_{2}{\mathrm{O}}_{6+\ensuremath{\delta}}$}},}\
  }\href {\doibase 10.1103/PhysRevB.71.094510} {\bibfield  {journal} {\bibinfo
  {journal} {Phys. Rev. B}\ }\textbf {\bibinfo {volume} {71}},\ \bibinfo
  {pages} {094510} (\bibinfo {year} {2005})}\BibitemShut {NoStop}%
\bibitem [{\citenamefont {Cava}\ \emph
  {et~al.}(1990{\natexlab{a}})\citenamefont {Cava}, \citenamefont {Batlogg},
  \citenamefont {van Dover}, \citenamefont {Krajewski}, \citenamefont
  {Waszczak}, \citenamefont {Fleming}, \citenamefont {Peck}, \citenamefont
  {Rupp}, \citenamefont {Marsh}, \citenamefont {James},\ and\ \citenamefont
  {Schneemeyer}}]{cava90b}%
  \BibitemOpen
  \bibfield  {author} {\bibinfo {author} {\bibfnamefont {R.~J.}\ \bibnamefont
  {Cava}}, \bibinfo {author} {\bibfnamefont {B.}~\bibnamefont {Batlogg}},
  \bibinfo {author} {\bibfnamefont {R.~B.}\ \bibnamefont {van Dover}}, \bibinfo
  {author} {\bibfnamefont {J.~J.}\ \bibnamefont {Krajewski}}, \bibinfo {author}
  {\bibfnamefont {J.~V.}\ \bibnamefont {Waszczak}}, \bibinfo {author}
  {\bibfnamefont {R.~M.}\ \bibnamefont {Fleming}}, \bibinfo {author}
  {\bibfnamefont {W.~F.}\ \bibnamefont {Peck}}, \bibinfo {author}
  {\bibfnamefont {L.~W.}\ \bibnamefont {Rupp}}, \bibinfo {author}
  {\bibfnamefont {P.}~\bibnamefont {Marsh}}, \bibinfo {author} {\bibfnamefont
  {A.~C. W.~P.}\ \bibnamefont {James}}, \ and\ \bibinfo {author} {\bibfnamefont
  {L.~F.}\ \bibnamefont {Schneemeyer}},\ }\bibfield  {title} {\enquote
  {\bibinfo {title} {{Superconductivity at 60 K in
  La$_{2-x}$Sr$_x$CaCu$_2$O$_6$: the simplest double-layer cuprate}},}\
  }\href@noop {} {\bibfield  {journal} {\bibinfo  {journal} {Nature}\ }\textbf
  {\bibinfo {volume} {345}},\ \bibinfo {pages} {602--604} (\bibinfo {year}
  {1990}{\natexlab{a}})}\BibitemShut {NoStop}%
\bibitem [{\citenamefont {Nguyen}\ \emph {et~al.}(1980)\citenamefont {Nguyen},
  \citenamefont {Er-Rakho}, \citenamefont {Michel}, \citenamefont {Choisnet},\
  and\ \citenamefont {Raveau}}]{nguy80}%
  \BibitemOpen
  \bibfield  {author} {\bibinfo {author} {\bibfnamefont {N.}~\bibnamefont
  {Nguyen}}, \bibinfo {author} {\bibfnamefont {L.}~\bibnamefont {Er-Rakho}},
  \bibinfo {author} {\bibfnamefont {C.}~\bibnamefont {Michel}}, \bibinfo
  {author} {\bibfnamefont {J.}~\bibnamefont {Choisnet}}, \ and\ \bibinfo
  {author} {\bibfnamefont {B.}~\bibnamefont {Raveau}},\ }\bibfield  {title}
  {\enquote {\bibinfo {title} {{Intercroissance de feuillets ``perovskites
  lacunaires'' et de feuillets type chlorure de sodium: Les oxydes
  La$_{2-x}$A$_{1+x}$Cu$_2$O$_{6-x/2}$ (A = Ca, Sr)}},}\ }\href {\doibase
  http://dx.doi.org/10.1016/0025-5408(80)90212-3} {\bibfield  {journal}
  {\bibinfo  {journal} {Mat. Res. Bull.}\ }\textbf {\bibinfo {volume} {15}},\
  \bibinfo {pages} {891--897} (\bibinfo {year} {1980})}\BibitemShut {NoStop}%
\bibitem [{\citenamefont {Bednorz}\ and\ \citenamefont
  {M\"uller}(1986)}]{bedn86}%
  \BibitemOpen
  \bibfield  {author} {\bibinfo {author} {\bibfnamefont {J.G.}\ \bibnamefont
  {Bednorz}}\ and\ \bibinfo {author} {\bibfnamefont {K.A.}\ \bibnamefont
  {M\"uller}},\ }\bibfield  {title} {\enquote {\bibinfo {title} {{Possible high
  $T_c$ superconductivity in the Ba--La--Cu--O system}},}\ }\href@noop {}
  {\bibfield  {journal} {\bibinfo  {journal} {Z. Phys. B}\ }\textbf {\bibinfo
  {volume} {64}},\ \bibinfo {pages} {189--193} (\bibinfo {year}
  {1986})}\BibitemShut {NoStop}%
\bibitem [{\citenamefont {Torrance}\ \emph {et~al.}(1988)\citenamefont
  {Torrance}, \citenamefont {Tokura}, \citenamefont {Nazzal},\ and\
  \citenamefont {Parkin}}]{torr88}%
  \BibitemOpen
  \bibfield  {author} {\bibinfo {author} {\bibfnamefont {J.~B.}\ \bibnamefont
  {Torrance}}, \bibinfo {author} {\bibfnamefont {Y.}~\bibnamefont {Tokura}},
  \bibinfo {author} {\bibfnamefont {A.}~\bibnamefont {Nazzal}}, \ and\ \bibinfo
  {author} {\bibfnamefont {S.~S.~P.}\ \bibnamefont {Parkin}},\ }\bibfield
  {title} {\enquote {\bibinfo {title} {{Metallic, but Not Superconducting,
  La-Ba (and La-Sr) Copper Oxides}},}\ }\href {\doibase
  10.1103/PhysRevLett.60.542} {\bibfield  {journal} {\bibinfo  {journal} {Phys.
  Rev. Lett.}\ }\textbf {\bibinfo {volume} {60}},\ \bibinfo {pages} {542--545}
  (\bibinfo {year} {1988})}\BibitemShut {NoStop}%
\bibitem [{\citenamefont {Izumi}\ \emph {et~al.}(1989)\citenamefont {Izumi},
  \citenamefont {Takayama-Muromachi}, \citenamefont {Nakai},\ and\
  \citenamefont {Asano}}]{izum89}%
  \BibitemOpen
  \bibfield  {author} {\bibinfo {author} {\bibfnamefont {F.}~\bibnamefont
  {Izumi}}, \bibinfo {author} {\bibfnamefont {E.}~\bibnamefont
  {Takayama-Muromachi}}, \bibinfo {author} {\bibfnamefont {Y.}~\bibnamefont
  {Nakai}}, \ and\ \bibinfo {author} {\bibfnamefont {H.}~\bibnamefont
  {Asano}},\ }\bibfield  {title} {\enquote {\bibinfo {title} {{Structure
  refinement of La$_{1.9}$Ca$_{1.1}$Cu$_2$O$_6$ with neutron powder diffraction
  data}},}\ }\href {\doibase https://doi.org/10.1016/0921-4534(89)90472-3}
  {\bibfield  {journal} {\bibinfo  {journal} {Physica C: Superconductivity}\
  }\textbf {\bibinfo {volume} {157}},\ \bibinfo {pages} {89--92} (\bibinfo
  {year} {1989})}\BibitemShut {NoStop}%
\bibitem [{\citenamefont {Caignaert}\ \emph {et~al.}(1990)\citenamefont
  {Caignaert}, \citenamefont {Nguyen},\ and\ \citenamefont {Raveau}}]{caig90}%
  \BibitemOpen
  \bibfield  {author} {\bibinfo {author} {\bibfnamefont {V.}~\bibnamefont
  {Caignaert}}, \bibinfo {author} {\bibfnamefont {N.}~\bibnamefont {Nguyen}}, \
  and\ \bibinfo {author} {\bibfnamefont {B.}~\bibnamefont {Raveau}},\
  }\bibfield  {title} {\enquote {\bibinfo {title} {{La$_2$SrCu$_2$O$_6$:
  Neutron diffraction study}},}\ }\href {\doibase
  http://dx.doi.org/10.1016/0025-5408(90)90046-5} {\bibfield  {journal}
  {\bibinfo  {journal} {Mat. Res. Bull.}\ }\textbf {\bibinfo {volume} {25}},\
  \bibinfo {pages} {199--204} (\bibinfo {year} {1990})}\BibitemShut {NoStop}%
\bibitem [{\citenamefont {Sakurai}\ \emph {et~al.}(1991)\citenamefont
  {Sakurai}, \citenamefont {Yamashita}, \citenamefont {Willis}, \citenamefont
  {Yamauchi}, \citenamefont {Tanaka},\ and\ \citenamefont {Kwei}}]{saku91}%
  \BibitemOpen
  \bibfield  {author} {\bibinfo {author} {\bibfnamefont {Takeshi}\ \bibnamefont
  {Sakurai}}, \bibinfo {author} {\bibfnamefont {Toru}\ \bibnamefont
  {Yamashita}}, \bibinfo {author} {\bibfnamefont {J.O.}\ \bibnamefont
  {Willis}}, \bibinfo {author} {\bibfnamefont {H.}~\bibnamefont {Yamauchi}},
  \bibinfo {author} {\bibfnamefont {Shoji}\ \bibnamefont {Tanaka}}, \ and\
  \bibinfo {author} {\bibfnamefont {George~H.}\ \bibnamefont {Kwei}},\
  }\bibfield  {title} {\enquote {\bibinfo {title} {{Combined X-ray and neutron
  powder diffraction study of the structure of
  La$_{1.8}$Sr$_{0.2}$CaCu$_2$O$_6$}},}\ }\href {\doibase
  http://dx.doi.org/10.1016/0921-4534(91)90435-2} {\bibfield  {journal}
  {\bibinfo  {journal} {Physica C: Superconductivity}\ }\textbf {\bibinfo
  {volume} {174}},\ \bibinfo {pages} {187--194} (\bibinfo {year}
  {1991})}\BibitemShut {NoStop}%
\bibitem [{\citenamefont {Liu}\ \emph {et~al.}(1991)\citenamefont {Liu},
  \citenamefont {Morris}, \citenamefont {Sinha},\ and\ \citenamefont
  {Tang}}]{liu91}%
  \BibitemOpen
  \bibfield  {author} {\bibinfo {author} {\bibfnamefont {H.~B.}\ \bibnamefont
  {Liu}}, \bibinfo {author} {\bibfnamefont {D.~E.}\ \bibnamefont {Morris}},
  \bibinfo {author} {\bibfnamefont {A.~P.~B.}\ \bibnamefont {Sinha}}, \ and\
  \bibinfo {author} {\bibfnamefont {X.~X.}\ \bibnamefont {Tang}},\ }\bibfield
  {title} {\enquote {\bibinfo {title} {{Superconductivity in
  La$_{2-x}$Sr$_x$CaCu$_2$O$_6$ at 60 K}},}\ }\href {\doibase
  http://dx.doi.org/10.1016/0921-4534(91)90417-W} {\bibfield  {journal}
  {\bibinfo  {journal} {Physica C: Superconductivity}\ }\textbf {\bibinfo
  {volume} {174}},\ \bibinfo {pages} {28--32} (\bibinfo {year}
  {1991})}\BibitemShut {NoStop}%
\bibitem [{\citenamefont {Fuertes}\ \emph {et~al.}(1990)\citenamefont
  {Fuertes}, \citenamefont {Obradors}, \citenamefont {Navarro}, \citenamefont
  {Gomez-Romero}, \citenamefont {Casa{\~n}-Pastor}, \citenamefont {P{\'e}rez},
  \citenamefont {Fontcuberta}, \citenamefont {Miravitlles}, \citenamefont
  {Rodriguez-Carvajal},\ and\ \citenamefont {Mart{\'\i}nez}}]{fuer90}%
  \BibitemOpen
  \bibfield  {author} {\bibinfo {author} {\bibfnamefont {A.}~\bibnamefont
  {Fuertes}}, \bibinfo {author} {\bibfnamefont {X.}~\bibnamefont {Obradors}},
  \bibinfo {author} {\bibfnamefont {J.M.}\ \bibnamefont {Navarro}}, \bibinfo
  {author} {\bibfnamefont {P.}~\bibnamefont {Gomez-Romero}}, \bibinfo {author}
  {\bibfnamefont {N.}~\bibnamefont {Casa{\~n}-Pastor}}, \bibinfo {author}
  {\bibfnamefont {F.}~\bibnamefont {P{\'e}rez}}, \bibinfo {author}
  {\bibfnamefont {J.}~\bibnamefont {Fontcuberta}}, \bibinfo {author}
  {\bibfnamefont {C.}~\bibnamefont {Miravitlles}}, \bibinfo {author}
  {\bibfnamefont {J.}~\bibnamefont {Rodriguez-Carvajal}}, \ and\ \bibinfo
  {author} {\bibfnamefont {B.}~\bibnamefont {Mart{\'\i}nez}},\ }\bibfield
  {title} {\enquote {\bibinfo {title} {{Oxygen excess and superconductivity at
  45 K in La$_2$CaCu$_2$O$_{6+y}$}},}\ }\href {\doibase
  http://dx.doi.org/10.1016/0921-4534(90)90241-6} {\bibfield  {journal}
  {\bibinfo  {journal} {Physica C: Superconductivity}\ }\textbf {\bibinfo
  {volume} {170}},\ \bibinfo {pages} {153--160} (\bibinfo {year}
  {1990})}\BibitemShut {NoStop}%
\bibitem [{\citenamefont {Kinoshita}\ \emph {et~al.}(1990)\citenamefont
  {Kinoshita}, \citenamefont {Shibata},\ and\ \citenamefont {Yamada}}]{kino90}%
  \BibitemOpen
  \bibfield  {author} {\bibinfo {author} {\bibfnamefont {Kyoichi}\ \bibnamefont
  {Kinoshita}}, \bibinfo {author} {\bibfnamefont {Hiroyuki}\ \bibnamefont
  {Shibata}}, \ and\ \bibinfo {author} {\bibfnamefont {Tomoaki}\ \bibnamefont
  {Yamada}},\ }\bibfield  {title} {\enquote {\bibinfo {title} {{High-pressure
  synthesis of superconducting
  La$_{2-x}$Ca$_{1+x}$Cu$_2$O$_{6-x/2+\delta}$}},}\ }\href {\doibase
  http://dx.doi.org/10.1016/0921-4534(90)90267-I} {\bibfield  {journal}
  {\bibinfo  {journal} {Physica C: Superconductivity}\ }\textbf {\bibinfo
  {volume} {171}},\ \bibinfo {pages} {523--527} (\bibinfo {year}
  {1990})}\BibitemShut {NoStop}%
\bibitem [{\citenamefont {Kinoshita}\ \emph {et~al.}(1992)\citenamefont
  {Kinoshita}, \citenamefont {Izumi}, \citenamefont {Yamada},\ and\
  \citenamefont {Asano}}]{kino92a}%
  \BibitemOpen
  \bibfield  {author} {\bibinfo {author} {\bibfnamefont {K.}~\bibnamefont
  {Kinoshita}}, \bibinfo {author} {\bibfnamefont {F.}~\bibnamefont {Izumi}},
  \bibinfo {author} {\bibfnamefont {T.}~\bibnamefont {Yamada}}, \ and\ \bibinfo
  {author} {\bibfnamefont {H.}~\bibnamefont {Asano}},\ }\bibfield  {title}
  {\enquote {\bibinfo {title} {{Structure refinements of superconducting and
  nonsuperconducting La$_{1.82}$Ca$_{1.18}$Cu$_2$O$_{6\pm\delta}$ from
  neutron-diffraction data}},}\ }\href {\doibase 10.1103/PhysRevB.45.5558}
  {\bibfield  {journal} {\bibinfo  {journal} {Phys. Rev. B}\ }\textbf {\bibinfo
  {volume} {45}},\ \bibinfo {pages} {5558--5562} (\bibinfo {year}
  {1992})}\BibitemShut {NoStop}%
\bibitem [{\citenamefont {Kinoshita}\ and\ \citenamefont
  {Yamada}(1992)}]{kino92b}%
  \BibitemOpen
  \bibfield  {author} {\bibinfo {author} {\bibfnamefont {K.}~\bibnamefont
  {Kinoshita}}\ and\ \bibinfo {author} {\bibfnamefont {T.}~\bibnamefont
  {Yamada}},\ }\bibfield  {title} {\enquote {\bibinfo {title}
  {{Superconductivity and antiferromagnetism in
  La$_{2-x}$Ca$_{1+x}$Cu$_2$O$_{6\pm\delta}$ and
  La$_{2-x}$Sr$_x$CaCu$_2$O$_{6\pm\delta}$}},}\ }\href {\doibase
  10.1103/PhysRevB.46.9116} {\bibfield  {journal} {\bibinfo  {journal} {Phys.
  Rev. B}\ }\textbf {\bibinfo {volume} {46}},\ \bibinfo {pages} {9116--9122}
  (\bibinfo {year} {1992})}\BibitemShut {NoStop}%
\bibitem [{\citenamefont {Ishii}\ \emph {et~al.}(1991)\citenamefont {Ishii},
  \citenamefont {Watanabe}, \citenamefont {Kinoshita},\ and\ \citenamefont
  {Matsuda}}]{ishi91}%
  \BibitemOpen
  \bibfield  {author} {\bibinfo {author} {\bibfnamefont {Takao}\ \bibnamefont
  {Ishii}}, \bibinfo {author} {\bibfnamefont {Takao}\ \bibnamefont {Watanabe}},
  \bibinfo {author} {\bibfnamefont {Kyoichi}\ \bibnamefont {Kinoshita}}, \ and\
  \bibinfo {author} {\bibfnamefont {Azusa}\ \bibnamefont {Matsuda}},\
  }\bibfield  {title} {\enquote {\bibinfo {title} {{Single crystal growth and
  superconductivity in La$_{1.87}$Ca$_{1.13}$Cu$_2$O$_6$}},}\ }\href {\doibase
  http://dx.doi.org/10.1016/0921-4534(91)90008-M} {\bibfield  {journal}
  {\bibinfo  {journal} {Physica C: Superconductivity}\ }\textbf {\bibinfo
  {volume} {179}},\ \bibinfo {pages} {39--42} (\bibinfo {year}
  {1991})}\BibitemShut {NoStop}%
\bibitem [{\citenamefont {Okuya}\ \emph {et~al.}(1994)\citenamefont {Okuya},
  \citenamefont {Kimura}, \citenamefont {Kobayashi}, \citenamefont {Shimoyama},
  \citenamefont {Kitazawa}, \citenamefont {Yamafuji}, \citenamefont {Kishio},
  \citenamefont {Kinoshita},\ and\ \citenamefont {Yamada}}]{okuy94}%
  \BibitemOpen
  \bibfield  {author} {\bibinfo {author} {\bibfnamefont {M.}~\bibnamefont
  {Okuya}}, \bibinfo {author} {\bibfnamefont {T.}~\bibnamefont {Kimura}},
  \bibinfo {author} {\bibfnamefont {R.}~\bibnamefont {Kobayashi}}, \bibinfo
  {author} {\bibfnamefont {J.}~\bibnamefont {Shimoyama}}, \bibinfo {author}
  {\bibfnamefont {K.}~\bibnamefont {Kitazawa}}, \bibinfo {author}
  {\bibfnamefont {K.}~\bibnamefont {Yamafuji}}, \bibinfo {author}
  {\bibfnamefont {K.}~\bibnamefont {Kishio}}, \bibinfo {author} {\bibfnamefont
  {K.}~\bibnamefont {Kinoshita}}, \ and\ \bibinfo {author} {\bibfnamefont
  {T.}~\bibnamefont {Yamada}},\ }\bibfield  {title} {\enquote {\bibinfo {title}
  {{Single-crystal growth and anisotropic electrical properties of
  (La$_{1-x}$Ca$_x$)$_2$CaCu$_2$O$_6$}},}\ }\href {\doibase 10.1007/BF00724560}
  {\bibfield  {journal} {\bibinfo  {journal} {J. Supercond.}\ }\textbf
  {\bibinfo {volume} {7}},\ \bibinfo {pages} {313--318} (\bibinfo {year}
  {1994})}\BibitemShut {NoStop}%
\bibitem [{\citenamefont {Gu}\ \emph {et~al.}(2006{\natexlab{a}})\citenamefont
  {Gu}, \citenamefont {Hucker}, \citenamefont {Kim}, \citenamefont {Tranquada},
  \citenamefont {Li},\ and\ \citenamefont {Moodenbaugh}}]{gu06b}%
  \BibitemOpen
  \bibfield  {author} {\bibinfo {author} {\bibfnamefont {G.D.}\ \bibnamefont
  {Gu}}, \bibinfo {author} {\bibfnamefont {M.}~\bibnamefont {Hucker}}, \bibinfo
  {author} {\bibfnamefont {Y.-J.}\ \bibnamefont {Kim}}, \bibinfo {author}
  {\bibfnamefont {J.M.}\ \bibnamefont {Tranquada}}, \bibinfo {author}
  {\bibfnamefont {Q.}~\bibnamefont {Li}}, \ and\ \bibinfo {author}
  {\bibfnamefont {A.R.}\ \bibnamefont {Moodenbaugh}},\ }\bibfield  {title}
  {\enquote {\bibinfo {title} {{Single-crystal growth and superconductivity of
  (La$_{1-x}$Sr$_x$)$_2$CaCu$_2$O$_{6+\delta}$}},}\ }\href {\doibase
  http://dx.doi.org/10.1016/j.jcrysgro.2005.11.026} {\bibfield  {journal}
  {\bibinfo  {journal} {J. Cryst. Growth}\ }\textbf {\bibinfo {volume} {287}},\
  \bibinfo {pages} {318--322} (\bibinfo {year}
  {2006}{\natexlab{a}})}\BibitemShut {NoStop}%
\bibitem [{\citenamefont {Gu}\ \emph {et~al.}(2006{\natexlab{b}})\citenamefont
  {Gu}, \citenamefont {H\"ucker}, \citenamefont {Kim}, \citenamefont
  {Tranquada}, \citenamefont {Dabkowska}, \citenamefont {Luke}, \citenamefont
  {Timusk}, \citenamefont {Gaulin}, \citenamefont {Li},\ and\ \citenamefont
  {Moodenbaugh}}]{gu06a}%
  \BibitemOpen
  \bibfield  {author} {\bibinfo {author} {\bibfnamefont {G.~D.}\ \bibnamefont
  {Gu}}, \bibinfo {author} {\bibfnamefont {M.}~\bibnamefont {H\"ucker}},
  \bibinfo {author} {\bibfnamefont {Y.-J.}\ \bibnamefont {Kim}}, \bibinfo
  {author} {\bibfnamefont {J.~M.}\ \bibnamefont {Tranquada}}, \bibinfo {author}
  {\bibfnamefont {H.}~\bibnamefont {Dabkowska}}, \bibinfo {author}
  {\bibfnamefont {G.~M.}\ \bibnamefont {Luke}}, \bibinfo {author}
  {\bibfnamefont {T.}~\bibnamefont {Timusk}}, \bibinfo {author} {\bibfnamefont
  {B.~D.}\ \bibnamefont {Gaulin}}, \bibinfo {author} {\bibfnamefont
  {Q.}~\bibnamefont {Li}}, \ and\ \bibinfo {author} {\bibfnamefont {A.~R.}\
  \bibnamefont {Moodenbaugh}},\ }\bibfield  {title} {\enquote {\bibinfo {title}
  {{Crystal growth and superconductivity of
  (La$_{1-x}$Ca$_x$)$_2$CaCu$_2$O$_{6+\delta}$}},}\ }\href {\doibase
  http://dx.doi.org/10.1016/j.jpcs.2005.10.153} {\bibfield  {journal} {\bibinfo
   {journal} {J. Phys. Chem. Solids}\ }\textbf {\bibinfo {volume} {67}},\
  \bibinfo {pages} {431--434} (\bibinfo {year}
  {2006}{\natexlab{b}})}\BibitemShut {NoStop}%
\bibitem [{Note1()}]{Note1}%
  \BibitemOpen
  \bibinfo {note} {{\protect \color {black}Note that the annealing temperature
  was selected to test the performance of the HIP. Typical annealing
  temperatures in previous studies have been somewhat lower, in the range of
  970$^\circ $C to 1080$^\circ $C \cite {cava90b,kino90,ishi91}.}}\BibitemShut
  {Stop}%
\bibitem [{\citenamefont {Stone}\ \emph {et~al.}(2014)\citenamefont {Stone},
  \citenamefont {Niedziela}, \citenamefont {Abernathy}, \citenamefont
  {DeBeer-Schmitt}, \citenamefont {Ehlers}, \citenamefont {Garlea},
  \citenamefont {Granroth}, \citenamefont {Graves-Brook}, \citenamefont
  {Kolesnikov}, \citenamefont {Podlesnyak},\ and\ \citenamefont
  {Winn}}]{tof_sns14}%
  \BibitemOpen
  \bibfield  {author} {\bibinfo {author} {\bibfnamefont {M.~B.}\ \bibnamefont
  {Stone}}, \bibinfo {author} {\bibfnamefont {J.~L.}\ \bibnamefont
  {Niedziela}}, \bibinfo {author} {\bibfnamefont {D.~L.}\ \bibnamefont
  {Abernathy}}, \bibinfo {author} {\bibfnamefont {L.}~\bibnamefont
  {DeBeer-Schmitt}}, \bibinfo {author} {\bibfnamefont {G.}~\bibnamefont
  {Ehlers}}, \bibinfo {author} {\bibfnamefont {O.}~\bibnamefont {Garlea}},
  \bibinfo {author} {\bibfnamefont {G.~E.}\ \bibnamefont {Granroth}}, \bibinfo
  {author} {\bibfnamefont {M.}~\bibnamefont {Graves-Brook}}, \bibinfo {author}
  {\bibfnamefont {A.~I.}\ \bibnamefont {Kolesnikov}}, \bibinfo {author}
  {\bibfnamefont {A.}~\bibnamefont {Podlesnyak}}, \ and\ \bibinfo {author}
  {\bibfnamefont {B.}~\bibnamefont {Winn}},\ }\bibfield  {title} {\enquote
  {\bibinfo {title} {{A comparison of four direct geometry time-of-flight
  spectrometers at the Spallation Neutron Source}},}\ }\href {\doibase
  10.1063/1.4870050} {\bibfield  {journal} {\bibinfo  {journal} {Rev. Sci.
  Instrum.}\ }\textbf {\bibinfo {volume} {85}},\ \bibinfo {pages} {045113}
  (\bibinfo {year} {2014})}\BibitemShut {NoStop}%
\bibitem [{\citenamefont {Suter}\ and\ \citenamefont {Wojek}(2012)}]{sute12}%
  \BibitemOpen
  \bibfield  {author} {\bibinfo {author} {\bibfnamefont {A.}~\bibnamefont
  {Suter}}\ and\ \bibinfo {author} {\bibfnamefont {B.M.}\ \bibnamefont
  {Wojek}},\ }\bibfield  {title} {\enquote {\bibinfo {title} {{{\tt musrfit}: A
  Free Platform-Independent Framework for μSR Data Analysis}},}\ }\href
  {\doibase https://doi.org/10.1016/j.phpro.2012.04.042} {\bibfield  {journal}
  {\bibinfo  {journal} {Physics Procedia}\ }\textbf {\bibinfo {volume} {30}},\
  \bibinfo {pages} {69--73} (\bibinfo {year} {2012})}\BibitemShut {NoStop}%
\bibitem [{\citenamefont {Sakurai}\ \emph {et~al.}(1993)\citenamefont
  {Sakurai}, \citenamefont {Yamashita},\ and\ \citenamefont
  {Yamauchi}}]{saku93}%
  \BibitemOpen
  \bibfield  {author} {\bibinfo {author} {\bibfnamefont {Takeshi}\ \bibnamefont
  {Sakurai}}, \bibinfo {author} {\bibfnamefont {Toru}\ \bibnamefont
  {Yamashita}}, \ and\ \bibinfo {author} {\bibfnamefont {H.}~\bibnamefont
  {Yamauchi}},\ }\bibfield  {title} {\enquote {\bibinfo {title} {{Stability of
  superconducting La$_{1.8}$Sr$_{0.2}$CaCu$_2$O$_6$ with respect to oxygen
  partial pressure}},}\ }\href {\doibase 10.1063/1.354094} {\bibfield
  {journal} {\bibinfo  {journal} {J. Appl. Phys.}\ }\textbf {\bibinfo {volume}
  {73}},\ \bibinfo {pages} {7575--7580} (\bibinfo {year} {1993})}\BibitemShut
  {NoStop}%
\bibitem [{\citenamefont {Hu}\ \emph {et~al.}(2014)\citenamefont {Hu},
  \citenamefont {Zhu}, \citenamefont {Shi}, \citenamefont {Li}, \citenamefont
  {Zhong}, \citenamefont {Schneeloch}, \citenamefont {Gu}, \citenamefont
  {Tranquada},\ and\ \citenamefont {Billinge}}]{hu14b}%
  \BibitemOpen
  \bibfield  {author} {\bibinfo {author} {\bibfnamefont {Hefei}\ \bibnamefont
  {Hu}}, \bibinfo {author} {\bibfnamefont {Yimei}\ \bibnamefont {Zhu}},
  \bibinfo {author} {\bibfnamefont {Xiaoya}\ \bibnamefont {Shi}}, \bibinfo
  {author} {\bibfnamefont {Qiang}\ \bibnamefont {Li}}, \bibinfo {author}
  {\bibfnamefont {Ruidan}\ \bibnamefont {Zhong}}, \bibinfo {author}
  {\bibfnamefont {John~A.}\ \bibnamefont {Schneeloch}}, \bibinfo {author}
  {\bibfnamefont {Genda}\ \bibnamefont {Gu}}, \bibinfo {author} {\bibfnamefont
  {John~M.}\ \bibnamefont {Tranquada}}, \ and\ \bibinfo {author} {\bibfnamefont
  {Simon J.~L.}\ \bibnamefont {Billinge}},\ }\bibfield  {title} {\enquote
  {\bibinfo {title} {{Nanoscale coherent intergrowthlike defects in a crystal
  of
  ${\mathrm{La}}_{1.9}{\mathrm{Ca}}_{1.1}{\mathrm{Cu}}_{2}{\mathrm{O}}_{6+\ensuremath{\delta}}$
  made superconducting by high-pressure oxygen annealing}},}\ }\href {\doibase
  10.1103/PhysRevB.90.134518} {\bibfield  {journal} {\bibinfo  {journal} {Phys.
  Rev. B}\ }\textbf {\bibinfo {volume} {90}},\ \bibinfo {pages} {134518}
  (\bibinfo {year} {2014})}\BibitemShut {NoStop}%
\bibitem [{\citenamefont {Jorgensen}\ \emph {et~al.}(1988)\citenamefont
  {Jorgensen}, \citenamefont {Dabrowski}, \citenamefont {Pei}, \citenamefont
  {Hinks}, \citenamefont {Soderholm}, \citenamefont {Morosin}, \citenamefont
  {Schirber}, \citenamefont {Venturini},\ and\ \citenamefont
  {Ginley}}]{jorg88}%
  \BibitemOpen
  \bibfield  {author} {\bibinfo {author} {\bibfnamefont {J.~D.}\ \bibnamefont
  {Jorgensen}}, \bibinfo {author} {\bibfnamefont {B.}~\bibnamefont
  {Dabrowski}}, \bibinfo {author} {\bibfnamefont {Shiyou}\ \bibnamefont {Pei}},
  \bibinfo {author} {\bibfnamefont {D.~G.}\ \bibnamefont {Hinks}}, \bibinfo
  {author} {\bibfnamefont {L.}~\bibnamefont {Soderholm}}, \bibinfo {author}
  {\bibfnamefont {B.}~\bibnamefont {Morosin}}, \bibinfo {author} {\bibfnamefont
  {J.~E.}\ \bibnamefont {Schirber}}, \bibinfo {author} {\bibfnamefont {E.~L.}\
  \bibnamefont {Venturini}}, \ and\ \bibinfo {author} {\bibfnamefont {D.~S.}\
  \bibnamefont {Ginley}},\ }\bibfield  {title} {\enquote {\bibinfo {title}
  {{Superconducting phase of
  ${\mathrm{La}}_{2}\mathrm{Cu}{\mathrm{O}}_{4+\ensuremath{\delta}}$: A
  superconducting composition resulting from phase separation}},}\ }\href
  {\doibase 10.1103/PhysRevB.38.11337} {\bibfield  {journal} {\bibinfo
  {journal} {Phys. Rev. B}\ }\textbf {\bibinfo {volume} {38}},\ \bibinfo
  {pages} {11337--11345} (\bibinfo {year} {1988})}\BibitemShut {NoStop}%
\bibitem [{\citenamefont {Moodenbaugh}\ \emph {et~al.}(1992)\citenamefont
  {Moodenbaugh}, \citenamefont {Sabatini}, \citenamefont {Xu}, \citenamefont
  {Ochab},\ and\ \citenamefont {Huber}}]{mood92}%
  \BibitemOpen
  \bibfield  {author} {\bibinfo {author} {\bibfnamefont {A.R.}\ \bibnamefont
  {Moodenbaugh}}, \bibinfo {author} {\bibfnamefont {R.L.}\ \bibnamefont
  {Sabatini}}, \bibinfo {author} {\bibfnamefont {Youwen}\ \bibnamefont {Xu}},
  \bibinfo {author} {\bibfnamefont {John}\ \bibnamefont {Ochab}}, \ and\
  \bibinfo {author} {\bibfnamefont {J.G.}\ \bibnamefont {Huber}},\ }\bibfield
  {title} {\enquote {\bibinfo {title} {{Solubility of Ca in superconducting
  La$_{2-x}$Ca$_x$CuO$_4$}},}\ }\href {\doibase
  http://dx.doi.org/10.1016/0921-4534(92)90272-E} {\bibfield  {journal}
  {\bibinfo  {journal} {Physica C: Superconductivity}\ }\textbf {\bibinfo
  {volume} {198}},\ \bibinfo {pages} {103--108} (\bibinfo {year}
  {1992})}\BibitemShut {NoStop}%
\bibitem [{\citenamefont {Er-Rakho}\ \emph {et~al.}(1988)\citenamefont
  {Er-Rakho}, \citenamefont {Michel},\ and\ \citenamefont {Raveau}}]{erra88}%
  \BibitemOpen
  \bibfield  {author} {\bibinfo {author} {\bibfnamefont {L.}~\bibnamefont
  {Er-Rakho}}, \bibinfo {author} {\bibfnamefont {C.}~\bibnamefont {Michel}}, \
  and\ \bibinfo {author} {\bibfnamefont {B.}~\bibnamefont {Raveau}},\
  }\bibfield  {title} {\enquote {\bibinfo {title}
  {{La$_{8-x}$Sr$_x$Cu$_8$O$_{20}$: An oxygen-deficient perovskite built of
  CuO$_6$, CuO$_5$, and CuO$_4$ polyhedra}},}\ }\href {\doibase
  http://dx.doi.org/10.1016/0022-4596(88)90138-7} {\bibfield  {journal}
  {\bibinfo  {journal} {J Solid State Chem.}\ }\textbf {\bibinfo {volume}
  {73}},\ \bibinfo {pages} {514--519} (\bibinfo {year} {1988})}\BibitemShut
  {NoStop}%
\bibitem [{\citenamefont {Tokura}\ \emph {et~al.}(1987)\citenamefont {Tokura},
  \citenamefont {Torrance}, \citenamefont {Nazzal}, \citenamefont {Huang},\
  and\ \citenamefont {Ortiz}}]{toku87}%
  \BibitemOpen
  \bibfield  {author} {\bibinfo {author} {\bibfnamefont {Y.}~\bibnamefont
  {Tokura}}, \bibinfo {author} {\bibfnamefont {J.~B.}\ \bibnamefont
  {Torrance}}, \bibinfo {author} {\bibfnamefont {A.~I.}\ \bibnamefont
  {Nazzal}}, \bibinfo {author} {\bibfnamefont {T.~C.}\ \bibnamefont {Huang}}, \
  and\ \bibinfo {author} {\bibfnamefont {C.}~\bibnamefont {Ortiz}},\ }\bibfield
   {title} {\enquote {\bibinfo {title} {{Discovery of a new, metallic (but not
  superconducting) compound in the lanthanum-strontium-copper-oxygen system:
  La$_5$SrCu$_6$O$_{15}$}},}\ }\href {\doibase 10.1021/ja00258a064} {\bibfield
  {journal} {\bibinfo  {journal} {J. Am. Chem. Soc.}\ }\textbf {\bibinfo
  {volume} {109}},\ \bibinfo {pages} {7555--7557} (\bibinfo {year}
  {1987})}\BibitemShut {NoStop}%
\bibitem [{\citenamefont {Borsa}\ \emph {et~al.}(1995)\citenamefont {Borsa},
  \citenamefont {Carretta}, \citenamefont {Cho}, \citenamefont {Chou},
  \citenamefont {Hu}, \citenamefont {Johnston}, \citenamefont {Lascialfari},
  \citenamefont {Torgeson}, \citenamefont {Gooding}, \citenamefont {Salem},\
  and\ \citenamefont {Vos}}]{bors95}%
  \BibitemOpen
  \bibfield  {author} {\bibinfo {author} {\bibfnamefont {F.}~\bibnamefont
  {Borsa}}, \bibinfo {author} {\bibfnamefont {P.}~\bibnamefont {Carretta}},
  \bibinfo {author} {\bibfnamefont {J.~H.}\ \bibnamefont {Cho}}, \bibinfo
  {author} {\bibfnamefont {F.~C.}\ \bibnamefont {Chou}}, \bibinfo {author}
  {\bibfnamefont {Q.}~\bibnamefont {Hu}}, \bibinfo {author} {\bibfnamefont
  {D.~C.}\ \bibnamefont {Johnston}}, \bibinfo {author} {\bibfnamefont
  {A.}~\bibnamefont {Lascialfari}}, \bibinfo {author} {\bibfnamefont {D.~R.}\
  \bibnamefont {Torgeson}}, \bibinfo {author} {\bibfnamefont {R.~J.}\
  \bibnamefont {Gooding}}, \bibinfo {author} {\bibfnamefont {N.~M.}\
  \bibnamefont {Salem}}, \ and\ \bibinfo {author} {\bibfnamefont {K.~J.~E.}\
  \bibnamefont {Vos}},\ }\bibfield  {title} {\enquote {\bibinfo {title}
  {{Staggered magnetization in
  ${\mathrm{La}}_{2\mathrm{\ensuremath{-}}\mathit{x}}$${\mathrm{Sr}}_{\mathit{x}}$${\mathrm{CuO}}_{4}$
  from $^{139}\mathrm{La}$ NQR and \ensuremath{\mu}SR: Effects of Sr doping in
  the range $0<x<0.02$}},}\ }\href {\doibase 10.1103/PhysRevB.52.7334}
  {\bibfield  {journal} {\bibinfo  {journal} {Phys. Rev. B}\ }\textbf {\bibinfo
  {volume} {52}},\ \bibinfo {pages} {7334--7345} (\bibinfo {year}
  {1995})}\BibitemShut {NoStop}%
\bibitem [{\citenamefont {Chou}\ \emph {et~al.}(1993)\citenamefont {Chou},
  \citenamefont {Borsa}, \citenamefont {Cho}, \citenamefont {Johnston},
  \citenamefont {Lascialfari}, \citenamefont {Torgeson},\ and\ \citenamefont
  {Ziolo}}]{chou93}%
  \BibitemOpen
  \bibfield  {author} {\bibinfo {author} {\bibfnamefont {F.~C.}\ \bibnamefont
  {Chou}}, \bibinfo {author} {\bibfnamefont {F.}~\bibnamefont {Borsa}},
  \bibinfo {author} {\bibfnamefont {J.~H.}\ \bibnamefont {Cho}}, \bibinfo
  {author} {\bibfnamefont {D.~C.}\ \bibnamefont {Johnston}}, \bibinfo {author}
  {\bibfnamefont {A.}~\bibnamefont {Lascialfari}}, \bibinfo {author}
  {\bibfnamefont {D.~R.}\ \bibnamefont {Torgeson}}, \ and\ \bibinfo {author}
  {\bibfnamefont {J.}~\bibnamefont {Ziolo}},\ }\bibfield  {title} {\enquote
  {\bibinfo {title} {{Magnetic phase diagram of lightly doped
  ${\mathrm{La}}_{2\mathrm{\ensuremath{-}}\mathit{x}}$${\mathrm{Sr}}_{\mathit{x}}$${\mathrm{CuO}}_{4}$
  from $^{139}\mathrm{La}$ nuclear quadrupole resonance}},}\ }\href {\doibase
  10.1103/PhysRevLett.71.2323} {\bibfield  {journal} {\bibinfo  {journal}
  {Phys. Rev. Lett.}\ }\textbf {\bibinfo {volume} {71}},\ \bibinfo {pages}
  {2323--2326} (\bibinfo {year} {1993})}\BibitemShut {NoStop}%
\bibitem [{\citenamefont {Matsuda}\ \emph {et~al.}(2002)\citenamefont
  {Matsuda}, \citenamefont {Fujita}, \citenamefont {Yamada}, \citenamefont
  {Birgeneau}, \citenamefont {Endoh},\ and\ \citenamefont {Shirane}}]{mats02}%
  \BibitemOpen
  \bibfield  {author} {\bibinfo {author} {\bibfnamefont {M.}~\bibnamefont
  {Matsuda}}, \bibinfo {author} {\bibfnamefont {M.}~\bibnamefont {Fujita}},
  \bibinfo {author} {\bibfnamefont {K.}~\bibnamefont {Yamada}}, \bibinfo
  {author} {\bibfnamefont {R.~J.}\ \bibnamefont {Birgeneau}}, \bibinfo {author}
  {\bibfnamefont {Y.}~\bibnamefont {Endoh}}, \ and\ \bibinfo {author}
  {\bibfnamefont {G.}~\bibnamefont {Shirane}},\ }\bibfield  {title} {\enquote
  {\bibinfo {title} {{Electronic phase separation in lightly doped
  ${\mathrm{La}}_{2\ensuremath{-}x}{\mathrm{Sr}}_{x}\mathrm{Cu}{\mathrm{O}}_{4}$}},}\
  }\href {\doibase 10.1103/PhysRevB.65.134515} {\bibfield  {journal} {\bibinfo
  {journal} {Phys. Rev. B}\ }\textbf {\bibinfo {volume} {65}},\ \bibinfo
  {pages} {134515} (\bibinfo {year} {2002})}\BibitemShut {NoStop}%
\bibitem [{\citenamefont {Niedermayer}\ \emph {et~al.}(1998)\citenamefont
  {Niedermayer}, \citenamefont {Bernhard}, \citenamefont {Blasius},
  \citenamefont {Golnik}, \citenamefont {Moodenbaugh},\ and\ \citenamefont
  {Budnick}}]{nied98}%
  \BibitemOpen
  \bibfield  {author} {\bibinfo {author} {\bibfnamefont {Ch.}\ \bibnamefont
  {Niedermayer}}, \bibinfo {author} {\bibfnamefont {C.}~\bibnamefont
  {Bernhard}}, \bibinfo {author} {\bibfnamefont {T.}~\bibnamefont {Blasius}},
  \bibinfo {author} {\bibfnamefont {A.}~\bibnamefont {Golnik}}, \bibinfo
  {author} {\bibfnamefont {A.}~\bibnamefont {Moodenbaugh}}, \ and\ \bibinfo
  {author} {\bibfnamefont {J.~I.}\ \bibnamefont {Budnick}},\ }\bibfield
  {title} {\enquote {\bibinfo {title} {{Common Phase Diagram for
  Antiferromagnetism in $\mathrm{La}_{2-x}\mathrm{Sr}_x\mathrm{CuO}_{4}$ and
  $\mathrm{Y}_{1-x}\mathrm{Ca}_x\mathrm{Ba}_2\mathrm{Cu}_3\mathrm{O}_6$ as Seen
  by Muon Spin Rotation}},}\ }\href {\doibase 10.1103/PhysRevLett.80.3843}
  {\bibfield  {journal} {\bibinfo  {journal} {Phys. Rev. Lett.}\ }\textbf
  {\bibinfo {volume} {80}},\ \bibinfo {pages} {3843--3846} (\bibinfo {year}
  {1998})}\BibitemShut {NoStop}%
\bibitem [{\citenamefont {Wells}\ \emph {et~al.}(1997)\citenamefont {Wells},
  \citenamefont {Lee}, \citenamefont {Kastner}, \citenamefont {Christianson},
  \citenamefont {Birgeneau}, \citenamefont {Yamada}, \citenamefont {Endoh},\
  and\ \citenamefont {Shirane}}]{well97}%
  \BibitemOpen
  \bibfield  {author} {\bibinfo {author} {\bibfnamefont {B.~O.}\ \bibnamefont
  {Wells}}, \bibinfo {author} {\bibfnamefont {Y.~S.}\ \bibnamefont {Lee}},
  \bibinfo {author} {\bibfnamefont {M.~A.}\ \bibnamefont {Kastner}}, \bibinfo
  {author} {\bibfnamefont {R.~J.}\ \bibnamefont {Christianson}}, \bibinfo
  {author} {\bibfnamefont {R.~J.}\ \bibnamefont {Birgeneau}}, \bibinfo {author}
  {\bibfnamefont {K.}~\bibnamefont {Yamada}}, \bibinfo {author} {\bibfnamefont
  {Y.}~\bibnamefont {Endoh}}, \ and\ \bibinfo {author} {\bibfnamefont
  {G.}~\bibnamefont {Shirane}},\ }\bibfield  {title} {\enquote {\bibinfo
  {title} {{Incommensurate Spin Fluctuations in High-Transition Temperature
  Superconductors}},}\ }\href {\doibase 10.1126/science.277.5329.1067}
  {\bibfield  {journal} {\bibinfo  {journal} {Science}\ }\textbf {\bibinfo
  {volume} {277}},\ \bibinfo {pages} {1067--1071} (\bibinfo {year}
  {1997})}\BibitemShut {NoStop}%
\bibitem [{\citenamefont {Dabrowski}\ \emph {et~al.}(1993)\citenamefont
  {Dabrowski}, \citenamefont {Wang}, \citenamefont {Jorgensen}, \citenamefont
  {Hitterman}, \citenamefont {Wagner}, \citenamefont {Hunter},\ and\
  \citenamefont {Hinks}}]{dabr93}%
  \BibitemOpen
  \bibfield  {author} {\bibinfo {author} {\bibfnamefont {B.}~\bibnamefont
  {Dabrowski}}, \bibinfo {author} {\bibfnamefont {Z.}~\bibnamefont {Wang}},
  \bibinfo {author} {\bibfnamefont {J.D.}\ \bibnamefont {Jorgensen}}, \bibinfo
  {author} {\bibfnamefont {R.L.}\ \bibnamefont {Hitterman}}, \bibinfo {author}
  {\bibfnamefont {J.L.}\ \bibnamefont {Wagner}}, \bibinfo {author}
  {\bibfnamefont {B.A.}\ \bibnamefont {Hunter}}, \ and\ \bibinfo {author}
  {\bibfnamefont {D.G.}\ \bibnamefont {Hinks}},\ }\bibfield  {title} {\enquote
  {\bibinfo {title} {{Suppression of superconducting transition temperature in
  orthorhombic La$_{2-x}$Ca$_x$CuO$_4$}},}\ }\href {\doibase
  https://doi.org/10.1016/0921-4534(93)90349-U} {\bibfield  {journal} {\bibinfo
   {journal} {Physica C: Superconductivity}\ }\textbf {\bibinfo {volume}
  {217}},\ \bibinfo {pages} {455--460} (\bibinfo {year} {1993})}\BibitemShut
  {NoStop}%
\bibitem [{\citenamefont {Blakeslee}\ \emph {et~al.}(1998)\citenamefont
  {Blakeslee}, \citenamefont {Birgeneau}, \citenamefont {Chou}, \citenamefont
  {Christianson}, \citenamefont {Kastner}, \citenamefont {Lee},\ and\
  \citenamefont {Wells}}]{blak98}%
  \BibitemOpen
  \bibfield  {author} {\bibinfo {author} {\bibfnamefont {P.}~\bibnamefont
  {Blakeslee}}, \bibinfo {author} {\bibfnamefont {R.~J.}\ \bibnamefont
  {Birgeneau}}, \bibinfo {author} {\bibfnamefont {F.~C.}\ \bibnamefont {Chou}},
  \bibinfo {author} {\bibfnamefont {R.}~\bibnamefont {Christianson}}, \bibinfo
  {author} {\bibfnamefont {M.~A.}\ \bibnamefont {Kastner}}, \bibinfo {author}
  {\bibfnamefont {Y.~S.}\ \bibnamefont {Lee}}, \ and\ \bibinfo {author}
  {\bibfnamefont {B.~O.}\ \bibnamefont {Wells}},\ }\bibfield  {title} {\enquote
  {\bibinfo {title} {{Electrochemistry and staging in
  ${\mathrm{La}}_{2}{\mathrm{CuO}}_{4+\mathrm{\ensuremath{\delta}}}$}},}\
  }\href {\doibase 10.1103/PhysRevB.57.13915} {\bibfield  {journal} {\bibinfo
  {journal} {Phys. Rev. B}\ }\textbf {\bibinfo {volume} {57}},\ \bibinfo
  {pages} {13915--13921} (\bibinfo {year} {1998})}\BibitemShut {NoStop}%
\bibitem [{\citenamefont {Radaelli}\ \emph {et~al.}(1993)\citenamefont
  {Radaelli}, \citenamefont {Jorgensen}, \citenamefont {Schultz}, \citenamefont
  {Hunter}, \citenamefont {Wagner}, \citenamefont {Chou},\ and\ \citenamefont
  {Johnston}}]{rada93}%
  \BibitemOpen
  \bibfield  {author} {\bibinfo {author} {\bibfnamefont {P.~G.}\ \bibnamefont
  {Radaelli}}, \bibinfo {author} {\bibfnamefont {J.~D.}\ \bibnamefont
  {Jorgensen}}, \bibinfo {author} {\bibfnamefont {A.~J.}\ \bibnamefont
  {Schultz}}, \bibinfo {author} {\bibfnamefont {B.~A.}\ \bibnamefont {Hunter}},
  \bibinfo {author} {\bibfnamefont {J.~L.}\ \bibnamefont {Wagner}}, \bibinfo
  {author} {\bibfnamefont {F.~C.}\ \bibnamefont {Chou}}, \ and\ \bibinfo
  {author} {\bibfnamefont {D.~C.}\ \bibnamefont {Johnston}},\ }\bibfield
  {title} {\enquote {\bibinfo {title} {{Structure of the superconducting
  ${\mathrm{La}}_{2}$${\mathrm{CuO}}_{4+\mathrm{\ensuremath{\delta}}}$ phases
  (${\delta}{\approx}0.08$, 0.12) prepared by electrochemical oxidation}},}\
  }\href {\doibase 10.1103/PhysRevB.48.499} {\bibfield  {journal} {\bibinfo
  {journal} {Phys. Rev. B}\ }\textbf {\bibinfo {volume} {48}},\ \bibinfo
  {pages} {499--510} (\bibinfo {year} {1993})}\BibitemShut {NoStop}%
\bibitem [{\citenamefont {Cava}\ \emph
  {et~al.}(1990{\natexlab{b}})\citenamefont {Cava}, \citenamefont {Santoro},
  \citenamefont {Krajewski}, \citenamefont {Fleming}, \citenamefont {Waszczak},
  \citenamefont {Peck},\ and\ \citenamefont {Marsh}}]{cava90c}%
  \BibitemOpen
  \bibfield  {author} {\bibinfo {author} {\bibfnamefont {R.J.}\ \bibnamefont
  {Cava}}, \bibinfo {author} {\bibfnamefont {A.}~\bibnamefont {Santoro}},
  \bibinfo {author} {\bibfnamefont {J.J.}\ \bibnamefont {Krajewski}}, \bibinfo
  {author} {\bibfnamefont {R.M.}\ \bibnamefont {Fleming}}, \bibinfo {author}
  {\bibfnamefont {J.V.}\ \bibnamefont {Waszczak}}, \bibinfo {author}
  {\bibfnamefont {W.F.}\ \bibnamefont {Peck}}, \ and\ \bibinfo {author}
  {\bibfnamefont {P.}~\bibnamefont {Marsh}},\ }\bibfield  {title} {\enquote
  {\bibinfo {title} {{The crystal structure of the
  La$_{1.6}$Sr$_{0.4}$CaCu$_2$O$_{6\pm\delta}$ superconductor}},}\ }\href
  {\doibase http://dx.doi.org/10.1016/0921-4534(90)90652-U} {\bibfield
  {journal} {\bibinfo  {journal} {Physica C: Superconductivity}\ }\textbf
  {\bibinfo {volume} {172}},\ \bibinfo {pages} {138--142} (\bibinfo {year}
  {1990}{\natexlab{b}})}\BibitemShut {NoStop}%
\bibitem [{\citenamefont {Shaked}\ \emph {et~al.}(1993)\citenamefont {Shaked},
  \citenamefont {Jorgensen}, \citenamefont {Hunter}, \citenamefont {Hitterman},
  \citenamefont {Kinoshita}, \citenamefont {Izumi},\ and\ \citenamefont
  {Kamiyama}}]{shak93}%
  \BibitemOpen
  \bibfield  {author} {\bibinfo {author} {\bibfnamefont {H.}~\bibnamefont
  {Shaked}}, \bibinfo {author} {\bibfnamefont {J.~D.}\ \bibnamefont
  {Jorgensen}}, \bibinfo {author} {\bibfnamefont {B.~A.}\ \bibnamefont
  {Hunter}}, \bibinfo {author} {\bibfnamefont {R.~L.}\ \bibnamefont
  {Hitterman}}, \bibinfo {author} {\bibfnamefont {K.}~\bibnamefont
  {Kinoshita}}, \bibinfo {author} {\bibfnamefont {F.}~\bibnamefont {Izumi}}, \
  and\ \bibinfo {author} {\bibfnamefont {T.}~\bibnamefont {Kamiyama}},\
  }\bibfield  {title} {\enquote {\bibinfo {title} {{Defect structure and
  superconducting properties of
  ${\mathrm{La}}_{1.8}$${\mathrm{Sr}}_{\mathit{x}}$${\mathrm{Ca}}_{1.2\mathrm{\ensuremath{-}}\mathit{x}}$${\mathrm{Cu}}_{2}$${\mathrm{O}}_{6\mathrm{\ensuremath{-}}\mathrm{\ensuremath{\delta}}}$}},}\
  }\href {\doibase 10.1103/PhysRevB.48.12941} {\bibfield  {journal} {\bibinfo
  {journal} {Phys. Rev. B}\ }\textbf {\bibinfo {volume} {48}},\ \bibinfo
  {pages} {12941--12950} (\bibinfo {year} {1993})}\BibitemShut {NoStop}%
\bibitem [{Note2()}]{Note2}%
  \BibitemOpen
  \bibinfo {note} {{\protect \color {black}In principle, one could estimate
  $x'$ from a refinement of neutron diffraction data; however, we did not
  collect data suitable for such a refinement. The triple-axis measurements
  covered only limited directions in reciprocal space. The time-of-flight
  measurements covered a bigger range of reciprocal space; however, those
  measurements were aimed at inelastic scattering and the data were not
  collected in a fashion that would ensure reliable Bragg peak
  intensities.}}\BibitemShut {Stop}%
\end{thebibliography}%

\end{document}